\documentclass[preprintnumbers,amsmath,amssymb,prd,superscriptaddress,notitlepage,nofootinbib] {revtex4-1}
\pdfoutput=1

\usepackage{graphicx}% Include figure files
\usepackage{dcolumn}% Align table columns on decimal point
\usepackage{bm}% bold math
\usepackage{subfigure}
\usepackage{multirow}
\usepackage{float} % force figures in page
\usepackage{ wasysym} 
\usepackage{color,ulem}

\newcommand{\frachalf}{\frac{1}{2}}

\newcommand{\Mpc}{\ \text{Mpc}}

\newcommand{\hMpc}{\ h^{-1}\text{Mpc}}

\newcommand{\ihMpc}{\ h\text{Mpc}^{-1}}

\newcommand{\be}{\begin{equation}}
\newcommand{\ee}{\end{equation}}

\newcommand{\Dtheta}{\Delta \theta}

\newcommand{\kms}{{\rm km~s^{-1}}}
\newcommand{\ikms}{{\rm s~km^{-1}}}

\newcommand{\sL}{\mathcal{L}}

\newcommand{\vtheta}{\mathbf{\theta}}
\newcommand{\vDtheta}{\mathbf{\Delta \theta}}

\newcommand{\vk}{\mathbf{k}}
\newcommand{\vq}{\mathbf{q}}

\newcommand{\vb}{\mathbf{b}}
\newcommand{\vx}{\mathbf{x}}

\newcommand{\vm}{\mathbf{m}}
\newcommand{\vd}{\mathbf{d}}
\newcommand{\vdelta}{\mathbf{\delta}}
\newcommand{\vy}{\mathbf{y}}

\newcommand{\vp}{\mathbf{p}}
\newcommand{\vep}{\mathbf{\epsilon}}

\newcommand{\vS}{\mathbf{S}}
\newcommand{\vC}{\mathbf{C}}

\newcommand{\vN}{\mathbf{N}}
\newcommand{\vA}{\mathbf{A}}
\newcommand{\vM}{\mathbf{M}}

\newcommand{\vT}{\mathbf{T}}
\newcommand{\vF}{\mathbf{F}}

\newcommand{\vR}{\mathbf{R}}

\newcommand{\kperp}{k_\perp}

\newcommand{\kpar}{k_\parallel}
\newcommand{\tk}{\tilde{k}}
\newcommand{\tmu}{\tilde{\mu}}
\newcommand{\PtD}{P_{\rm 3D}}
\newcommand{\PtDp}{P_{\rm 3D^\prime}}

\newcommand{\vL}{\mathbf{L}}

\newcommand{\vD}{\mathbf{D}}

\newcommand{\lya}{Ly$\alpha$}

\newcommand{\lyaf}{Ly$\alpha$ forest}

\def\pvmhid#1{}

\def\afhid#1{}

\def\ashid#1{}

\newcommand{\px}{p^\times}
\newcommand{\ptD}{p^{\rm 3D}}

\begin{document}
\title{How to estimate the 3D power spectrum of the Lyman-$\alpha$ forest}

\author{Andreu Font-Ribera \footnote{author list alphabetized}}
\email{a.font@ucl.ac.uk} 
\affiliation{Department of Physics and Astronomy, University College London, 
Gower Street, London, United Kingdom}
\author{Patrick McDonald}
\email{PVMcDonald@lbl.gov} 
\affiliation{Lawrence Berkeley National Laboratory, One Cyclotron Road,
Berkeley, CA 94720, USA}
\author{An\v{z}e Slosar}
\email{anze@bnl.gov} 
\affiliation{Brookhaven National Laboratory, Upton, NY 11973, USA}

\date{\today}

\begin{abstract}
  We derive and numerically implement an algorithm for estimating the
  3D power spectrum of the Lyman-$\alpha$ (\lya) forest flux fluctuations. 
  The algorithm  exploits the unique geometry of \lyaf\ data to efficiently 
  measure the cross-spectrum between lines of sight as a function of parallel
  wavenumber, transverse separation and redshift.
  We start by approximating the global covariance matrix as block-diagonal, 
  where only pixels from the same spectrum are correlated.
  We then compute the eigenvectors of the derivative of the signal covariance 
  with respect to cross-spectrum parameters, and project the 
  inverse-covariance-weighted spectra onto them. 
  This acts much like a radial Fourier transform over redshift windows. 
  The resulting cross-spectrum inference is then converted into our final
  product, an approximation of the likelihood for the 3D
  power spectrum expressed as second order Taylor expansion
  around a fiducial model. We demonstrate the accuracy and
  scalability of the algorithm and comment on possible extensions. Our
  algorithm will allow efficient analysis of the upcoming Dark Energy
  Spectroscopic Instrument dataset.
\end{abstract}

\maketitle

\section{Introduction}
\label{sec:intro}

Inference of the power spectra of cosmic fields is one of the main
tools of observational cosmology. Theory behind estimation of power
spectra for Cosmic Microwave Background and galaxy surveys (and other
point tracers) has been well developed over the past thirty years.  It
is not difficult to write a likelihood for the power spectrum given
measured fields assuming these to be Gaussian. From this one can
derive the standard ``optimal quadratic estimator''\footnote{Strictly 
speaking, the OQE is only approximately
quadratic and optimal, even for Gaussian fields, although in the large data 
set limit the distinction
is not important. The 
OQE equation as usually written appears at first glance to be quadratic in 
the data, but
it only achieves a maximum likelihood solution after iteration, 
which makes the final answer non-quadratic in
the input data (it is easy to see this by noting that the Fisher 
matrix,
i.e., 2nd derivative of the likelihood, appears in the standard formula, while
the maximum likelihood solution can only require the 1st derivative).
} (OQE) 
\cite{1998PhRvD..57.2117B,1998ApJ...503..492S,2000ApJ...533...19B},
but it is numerically too expensive to evaluate for
large surveys using brute-force linear algebra. Therefore, a number of
good approximations have been developed
\cite{1994ApJ...426...23F,1998ApJ...499..555T,2006PhRvD..74l3507T}.

In the field of \lyaf, the situation is much less developed.  
\cite{2006ApJS..163...80M} used an OQE to measure the \lyaf\ one-dimensional 
(1D) power spectrum, describing the correlations between pixels in a given 
line of sight, and a similar algorithm was more recently used in 
\cite{2013A&A...559A..85P}.
However, as the density of lines of sight increases in present and future
\lyaf\ surveys, it becomes more important to measure the full 
three-dimensional (3D) correlations, including correlations between pixels
from different lines of sight.
Recent analyses of 3D correlations, focused on measuring the large scale
feature of Baryon Acoustic Oscillation (BAO), have relied on measuring 
the correlation function using simple pixel-product algorithms
\cite{2011JCAP...09..001S,2013A&A...552A..96B,2015A&A...574A..59D,
2017A&A...603A..12B} (although see \cite{2013JCAP...04..026S} for an example 
of using OQE-like approach to estimate the correlation function).
The main goal of this paper is to present a likelihood based algorithm that 
will allow us to measure the 3D power spectrum of the \lyaf\ down to 
small scales, and including all the information in traditional 1D analyses. 

In principle the power spectrum and correlation function are simply
linear transforms of one another, potentially containing the same
information. In spite of this,
in practice they are often thought of as living in disconnected
worlds, i.e., estimated using much different algorithms and not
compared directly. Each has potential advantages for isolating
different kinds of systematics into a limited range of their
coordinates (e.g., the power spectrum is good for isolating slowly
varying effects to small wavenumbers, while the correlation function
is good for isolating effects at specific
separations). Different tricks and approximations are generally used
to make estimation algorithms tractable for large data sets
(including, most importantly, estimation of error bars). The power
spectrum has the advantage of generally less correlated errors and
diagonalization of the theory in the linear regime limit.  The primary
point of this paper is to provide a practical algorithm for the 3D
\lyaf\ power spectrum measurement, in order to exploit its advantages,
but an implicit point is that the different statistics do not need to live in 
isolated
worlds. We show in passing how a measurement of the power spectrum can
always be converted into a correlation function measurement and vice
versa, so that ultimately one should be able to exploit the advantages
of both within the same basic analysis (our conversion of the
cross-spectrum to power spectrum is an explicit example of this).

Methods developed for point tracers do not adapt directly to the
\lyaf\ problem, because of the unique window function of the \lyaf. We
measure the flux fields in individual lines of sight, each of which
covers a certain redshift range, which do not overlap perfectly and
which generally offer relatively high resolution in the radial direction but
only sparse sampling in the transverse direction.

This work is clearly timely given the forthcoming large increases in
the density of lines of sight in upcoming \lyaf\ surveys such as 
Dark Energy Spectroscopic Instrument (DESI, \cite{2013arXiv1308.0847L}).
The analysis of these datasets will present numerous challenges beyond power 
spectrum estimation in terms a plethora of observational and instrumental
systematics. In the current work we sweep these under the rug and
assume ideal measurements. However, ultimately one of the purposes of our 
general framework is to allow
understanding and propagation of these effects.  

We expect that applying Fast Fourier Transform (FFT) algorithm to each
spectrum will lead to algorithmic improvements, since each line
of sight's correlations are approximately stationary. This also
performs ``scale sorting'', which isolates large-scale radial 
modes that correlate across large transverse separations from small scale modes
which have significant correlations across considerably smaller
transverse separations. This line of thought naturally brings
cross-spectrum, $P_\times(z,\Dtheta,\kpar)$, between lines of
sight as a function of angular separation as a natural intermediate
product,\footnote{In this paper we follow the jargon of 
\cite{1999ApJ...511L...5H}
in calling the radial Fourier transform of the 3D \lyaf\ flux correlation 
function at non-zero transverse separation 
the ``cross-spectrum.'' This is not to be confused with the  
three-dimensional cross-power spectrum between
multiple tracers, e.g. the 3D Fourier transform of the cross-correlation 
function between quasar positions and \lyaf.} from which
the 3D power spectrum can be later inferred.  In this paper we bring
this rough idea to completion, by presenting a detailed algorithm and
its numerical implementation. We begin with the likelihood function
and then develop the numerically efficient and mathematically rigorous
derivation of expressions for cross-spectrum with well
understood approximations and then demonstrate how this can be
transformed into 3D power spectrum. The final product will be a
likelihood for 3D power spectrum as functions of redshift and of
radial and transverse wavenumbers.

We start in section \ref{sec:Like} by presenting the basic likelihood analysis
formalism we follow in this paper, and the specific case of a clustering 
analysis of the \lyaf\ is discussed in section \ref{sec:Lya}.
The numerical implementation is detailed in section \ref{sec:numeric}.
In section \ref{sec:P1D} we revisit 1D analyses in this context, and we show 
examples of analyses on mock datasets.
Measurements of the cross-power spectrum on mock datasets are presented in
\ref{sec:PX}, and in section \ref{sec:P3D} we describe how these can be 
translated into measurements of the 3D power spectrum.
We conclude with a discussion in section \ref{sec:discussion}.

\section{Likelihood approximation vs. standard power spectrum 
``estimators'' \label{sec:Like}}

While our algorithm should obtain results in the end essentially equivalent 
to the standard OQE, i.e., it could be 
understood simply as
a computing optimization, we think it is useful to take a step back
and shed some of the historical baggage that comes along with the
words in OQE. 
In cosmological applications, the OQE
has been introduced first into analysis of the CMB and galaxy survey
data
\cite{1997PhRvD..55.5895T,1998PhRvD..57.2117B,1998ApJ...499..555T,
2000ApJ...533...19B}. There
are two canonical derivations that appear in these papers. In the
first one, a general quadratic estimator is written, the weights are
then chosen to ensure inverse variance weighting and unbiased
estimates and then Cramer-Rao inequality is used to demonstrate the
estimator to be optimal (however, strictly speaking this demonstration 
requires knowing the correct answer in advance to use for the weighting). 
In the second derivation, the likelihood is
written and then the estimator is presented as a Newton-Raphson step
towards maximum likelihood with second derivative replaced by Fisher
matrix for numerical efficiency. These are both valid approaches, but
the OQE process is commonly viewed as simply an estimator that
transforms data into summary statistics. In this paper, we want to
connect the same fundamental mathematics to a more Bayesian thinking
about the likelihood. 
In particular, we believe that understanding the measurement 
process explicitly as a likelihood expansion leads to a streamlined 
understanding
of how we can convert between parameter bases, e.g., between cross-spectrum,
correlation function, and power spectrum parameters.  

Standard Bayesian data analysis says that for any data set
$\vd$ and parameters $\vp$ we should compute the posterior probability 
for the parameters given the data, $L(\vp| \vd)$.
Generally we do
this using Bayes theorem, $L(\vp| \vd) \propto L(\vd| \vp) L(\vp)$
where $L(\vp)$ is a prior on the parameters. In multi-dimensional
parameter spaces it is frequently not possible to fully tabulate
$L(\vp| \vd)$, let alone obtain an analytic form, so we generally need
to look for some limited approximate representation of it, sufficient
for the use we intend.  One such representation is a set of samples of
the parameter vector -- instances of $\vp$ distributed
consistent with $L(\vp| \vd)$, e.g., from MCMC. These are convenient
if one wants to integrate over $L(\vp| \vd)$, e.g., to produce means
or variances or confidence intervals for $\vp$. 

An alternative is to approximate $\sL\equiv \ln L$ by a Taylor series 
around $\vp_0$, i.e.,
$\sL(\vd|\vp) = \sL(\vd|\vp_0)+\sL_{,\alpha}(\vd|\vp_0) \delta p_\alpha +
\frachalf \sL_{,\alpha\beta}(\vd|\vp_0) \delta p_\alpha \delta p_\beta+...$,
where $\delta \vp\equiv \vp-\vp_0$. If we truncate this series at
quadratic order this is equivalent to the ubiquitous assumption that
the errors are Gaussian. 
This assumption is exact if
likelihood has a Gaussian shape and the theory is linear in
parameters, but it is in general a good approximation due to central
limit theorem or if theory is effectively linear over the range of
high-likelihood region. In the power spectrum measurement problem it
is wrong in sample-variance dominated measurement when only few
modes are present (e.g. the low-$\ell$ CMB power spectrum).
A key point here, especially relevant to
band power parameters, is if we have this Taylor series expansion
around parameters $\vp_0$ close enough to the values where we want to
use the likelihood, {\it we are done}, i.e., there is no fundamental
need to estimate ``best'' values for the parameters, as this is
irrelevant to further use of the likelihood function.  In fact, for
band power parameters where we expect the true power to be smooth, 
it generally makes
more sense to expand around our best {\it global} estimate of parameter
values, e.g., their values in our favorite cosmological model
constrained by all available datasets. Moreover, for the same reason it is
better not to perform ``iterative'' expansion around maximum
likelihood, because the covariance matrix given by an expansion 
around the best
guess of \emph{true} cosmological parameter is going to be more
accurate than that from an expansion around a model that is the best 
fit to only one of the datasets (for a
concrete example, see \cite{2004MNRAS.348..885E}). This does not
preclude identifying deviations from the best-guess model, as the likelihood
expansion still implies some maximum likelihood parameter values -- they
just are not necessarily $\vp_0$. At most, iteration should take place at the
most global possible level, e.g., possibly adjusting $\vp_0$ if the results
of all-data cosmological parameter estimation give a change in the best model.\footnote{In contrast, the most standard
``estimate the parameter values and covariance matrix'' form of data
analysis is to find maximum likelihood values of parameters, where
$\sL_{,\alpha}(\vp_0)=0$, and expand around this point to find the
covariance matrix.}

Given $\sL_{,\alpha}(\vp_0)$ and $\sL_{,\alpha\beta}(\vp_0)$, finding the
constraints on a different set of parameters, centered on the same fiducial
model and assuming a Taylor series for $\sL$ is still sufficient, is generally 
as simple as applying the derivative chain rule, e.g., 
$\frac{\partial\sL}{\partial\vp}= 
\frac{\partial\vq}{\partial\vp}\frac{\partial\sL}{\partial\vq}$, 
where $\vq$ is the new set of parameters.\footnote{As we discuss below, minor subtleties do arise in the
definition of $\partial\vq/\partial\vp$ for band power measurements.}

To make this more concrete for our problem: while the goal of a power spectrum 
measurement is often presented as being to
produce a set of estimated values for band parameters and their covariance
matrix (e.g., the $P_{1D}(z,\kpar)$ estimates of
\cite{2006ApJS..163...80M} and \cite{2013A&A...559A..85P}),
we consider our task to be to determine the coefficients $\sL_{,\alpha}$ and
$\sL_{,\alpha\beta}$.

We describe our data as $\vd=\vm+\vR\vdelta+\vep$, 
where $\vm=\left<\vd\right>$ is the mean, $\vR$ is the response matrix, 
$\vdelta$ is the cosmological fluctuation of interest and $\vep$ is the 
instrumental noise. We assume
a Gaussian likelihood for the data given a model,
\begin{equation}
L(\vd|\vp) \propto
  \det\left(\vC\right)^{-1/2}
  \exp\left[ -\frac{1}{2}\left(\vd-\vm\right)^t \vC^{-1} 
      \left(\vd-\vm\right) \right] ~,
\end{equation}
where $\vC=\vR\vS\vR^t+\vN$ with $\vS$ the covariance of $\vdelta$ (the
signal we want to measure) and 
$\vN$ the (diagonal) noise matrix. 
In the next section we will discuss in more detail what $\vm$ and $\vR$ are 
in the context of \lyaf\ analyses, but what matters here is that they do not
depend on the parameters $\vp$ we are trying to estimate.

As usual in power spectrum estimation, we assume that $\vS$ is linear in some 
parameters.
The derivatives we need are now quite simple:
\begin{equation}
\sL_{,\alpha}=\frac{1}{2} \left(\vd-\vm\right)^t 
  \vC_0^{-1} \vC_{,\alpha} \vC_0^{-1} \left(\vd-\vm\right)
  - \frac{1}{2}{\rm Tr}\left[\vC_0^{-1} \vC_{,\alpha}\right]~,
\label{eq:dL1}
\end{equation}
and
\begin{equation}
\sL_{,\alpha\beta}=
- \left(\vd-\vm\right)^t \vC_0^{-1} \vC_{,\alpha} \vC_0^{-1} 
  \vC_{,\beta} \vC_0^{-1} \left(\vd-\vm\right)
+ \frac{1}{2} {\rm Tr} \left[\vC_0^{-1} \vC_{,\alpha} \vC_0^{-1} \vC_{,\beta}\right]~,
\label{eq:dL2}
\end{equation}
where $\vC_0$ is the covariance matrix evaluated using $\vp_0$, and 
$\vC_{,\alpha}$ is the derivative of the covariance matrix with respect to the
parameter $p_\alpha$.
These Taylor series coefficients imply some mean (maximum likelihood)
values for $\vp$, and a covariance
matrix around that maximum, 
but those quantities are superfluous -- we are generally going to
use our power spectrum results to constrain some more fundamental parameters
and for that we just need $L(\vd | \vp)$.

The second derivatives have two terms: the first involves the measured data
$\vd$, and the second is a trace of a matrix product.
The expectation value of the first term is proportional to the second term 
if $\vC_0$ is the correct covariance matrix for the data, so we approximate 
the second derivative by its expected value under this assumption, as is 
usually done in deriving the OQE:
\begin{equation}
\sL_{,\alpha\beta} \approx
  \left< \sL_{,\alpha\beta} \right>_0 = - \frac{1}{2} {\rm Tr}
  \left[\vC_0^{-1} \vC_{,\alpha} \vC_0^{-1} \vC_{,\beta}\right]
  = - F_{\alpha\beta} ~,
\label{eq:Fisher}
\end{equation}
where $F_{\alpha\beta}$ is the Fisher matrix. 
In general we do not need to make this approximation and it may not actually 
be helpful, but we do not investigate it further in this paper
(see \cite{2015JCAP...01..022M} for a discussion about this approximation). 

The traditional OQE is derived as
a Newton-Raphson (NR) step toward the maximum 
likelihood
value of parameters, including this approximation for the 2nd derivative 
matrix, i.e., the standard OQE equation is simply a re-shuffling of terms in:
\begin{equation}
 \hat \vp = \vp_0 + \vF^{-1} \sL^\prime ~,
 \label{eq:NR}
\end{equation}
where $\vF$ is the Fisher matrix and $\sL^\prime$ the first derivatives of the
log-likelihood. 

Our bottom line point of this section is that in this paper we do not consider
this NR step to be a fundamental part of our analysis. 
Our purpose is
to estimate the likelihood derivatives, and we only compute things like the
implied central value when necessary for diagnostics like plotting. 

In this paper we often ignore the priors on the parameters, since they only 
play two very minor roles: i) we use them to compute implied central values
and errorbars when making plots; ii) in section \ref{sec:P3D} we marginalize
over some of the band powers, and we need to take into account the priors on 
the parameters that we marginalize out (see appendix \ref{sec:margin} on 
marginalization). 
Whenever they are required, we use Gaussian priors around zero, with a 
width set to a thousand times the fiducial power evaluated at the center of 
the band.\footnote{We have tested that the exact widths of the priors do not 
qualitatively change the results of this paper.}

\section{Clustering analyses of the \lyaf}
\label{sec:Lya}

In this section we will introduce basic notation in analyses of the \lyaf, 
present our data model and discuss the parameterization of the power spectrum.

\subsection{Data vector}

In a clustering analysis of the \lyaf\ the data set consists of a set of 
optical spectra of high redshift quasars, identified by an angular position 
$\vtheta$ and a redshift $z_q$. 
In the analysis of a real survey we would need to discuss instrumental 
specifications and details of the data reduction pipeline: sky subtraction, 
spectral calibration, co-addition of individual exposures, estimates of the 
noise variance, etc. 
In this study we will not address these issues in detail, although we will 
discuss how they would affect our measurements in section \ref{sec:discussion}.

We will use a simplified data model to describe the measured flux $d_i$ in a
pixel of observed wavelength $\lambda_i$ in the spectrum of quasar $q$ as:
\begin{equation}
 d_i = \int d\lambda ~ W_i(\lambda) ~ C_q(\lambda) ~ F(\lambda) + \epsilon_i ~,
\end{equation}
where $C_q(\lambda)$ is the quasar continuum, 
$F(\lambda) = e^{-\tau(\lambda)}$ is the transmitted flux fraction, 
$\tau(\lambda)$ the optical depth and $\epsilon_i$ the instrumental noise 
in the pixel.
$W_i(\lambda)$ is the convolution kernel for pixel $i$, including both spectral 
resolution and the pixelization (we will discuss this in detail soon).

In cosmological analyses of the \lyaf\ we are interested in measuring the 
statistics of the fluctuations around the mean transmitted flux fraction 
(often referred as the \textit{mean flux}), 
$F(\lambda) = \bar F(\lambda) \left(1 + \delta_F(\lambda) \right)$,
giving, in our simplified model,
\begin{equation}
 d_i = m_i  \left(1+\int d\lambda ~ W_i(\lambda) ~ \delta(\lambda)
 \right) + \epsilon_i ~,
 \label{eq:data_vector}
\end{equation}
where we have assumed that $C(\lambda) \bar F(\lambda)$ can be approximated 
as constant over the width of the smoothing kernel, and defined 
$m_i = C(\lambda_i) \bar F(\lambda_i)$.

\subsection{Covariances}

The covariance of the observed data vector (defined in
Eq. \ref{eq:data_vector}) is given by:
\begin{equation}
 C_{ij} = \left< d_i ~ d_j \right> - \left< d_i \right> \left< d_j \right> =
  m_i ~ m_j \int d\lambda ~ W_i(\lambda) 
  \int d\lambda^\prime ~ W_j(\lambda^\prime)
  ~ \xi_{3D}(\Dtheta_{ij},\lambda,\lambda^\prime)
  + N_{ij} ~,
 \label{eq:covar}
\end{equation}
where $N_{ij} = \sigma^2_i \delta_{ij}^K$ is the (diagonal) noise matrix
and $\xi_{3D}(\Dtheta_{ij},\lambda,\lambda^\prime)$ is the 3D correlation
function of the $\delta$ field we are interested in, and $\Dtheta_{ij}$ is 
the angular separation between the spectra of pixels $i$ and $j$.
Defining $\Delta v_{ij} = c \left( \ln{\lambda_j} - \ln{\lambda_i} \right)$, 
we can exactly equivalently describe
the 3D correlations as a function of angular separation, velocity separation 
and redshift, $\xi_{3D}(z_{ij},\Dtheta_{ij},\Delta v_{ij})$, with 
$1+z_{ij} = \sqrt{(1+z_i)(1+z_j)} = \sqrt{\lambda_i\lambda_j}/\lambda_\alpha$.
Note that this is equivalent to defining
$\bar{v}_{ij}\equiv c \left( \ln{\lambda_j} + \ln{\lambda_i} \right)/2$
and then $1+z_{ij}\equiv\exp\left(\bar{v}_{ij}/c\right)/\lambda_\alpha$.
These quantities should be understood as just different ways
of packaging the two observable wavelengths -- we do not rely on them having
any more fundamental physical meaning.

In the presence of redshift evolution, the power spectrum is not 
well-defined as a variance of Fourier modes. We can, however, still model the
data as a Gaussian random field, understood to mean a field with Gaussian 
likelihood function specified by a correlation function, which does not 
require the field to be stationary.  
We {\it define} the 3D power spectrum to be the Fourier transform of this 
correlation function:
\begin{equation}
 P_{3D}(z,k_\perp,k_\parallel) \equiv
  \int d \vDtheta ~ e^{i \vDtheta \vk_\perp}
  \int d \Delta v ~ e^{i \Delta v k_\parallel} 
  ~ \xi_{3D}(z,\Dtheta,\Delta v) ~,
\label{eq:P3Dfromxi}
\end{equation}
where $k_\perp$ has units of inverse radian, and $k_\parallel$ units of 
$\ikms$. We expect that on scales smaller than the scale of the evolution,
the power spectrum defined this way will closely follow intuition based on
the ideal stationary case, but in any case it is well-defined without any 
assumption about that. 
Note that it is not 
completely trivial
here that the Fourier transform is defined to be with respect to observable 
coordinates, rather than, e.g., an approximation of comoving coordinates using
a fiducial model. One should not think, however, 
that the BAO feature is directly smeared by this definition 
-- a delta function at some separation in the 
correlation function in comoving
coordinates remains a delta function in observable coordinates.\footnote{The key is to remember that by definition 
we must consider pairs of pixels at fixed $z_{ij}$ -- there is no averaging
together here of, e.g., purely transverse pixel pairs at different redshifts 
where fixed angular separation means different comoving separation.}

Given a comoving 
coordinate power spectrum we can always compute this observable coordinate
power spectrum by transforming to a correlation function, introducing the 
observable coordinates, and transforming back. Note, however, that on the
BAO scale the effect of ignoring the non-linearity of the coordinate transform 
is tiny.  E.g., in a typical
model the radial comoving separation between $z=2.31$ and $z=2.39$ is 
68.358 Mpc/h, while $(0.08/3.35) c H(z=2.35)=68.400$ Mpc/h, 
a $\sim 0.06$\% difference. 
If points at the same two redshifts are separated by 0.02 radians, the true
comoving transverse separation is 78.418 Mpc/h, while 0.02 times the angular 
diameter distance at $z=2.35$ is 78.425 Mpc/h, $\sim 0.009$\% different. I.e.,
at foreseeable levels of precision we can consider the relation between power
spectra to be a simple linear change of units, except possibly on much larger
scales. The key is that we keep separation calculations symmetric around the
central redshift, so that differences are quadratic in the 
scale changes 
(strictly speaking we measure distances symmetrically in $\ln(1+z)$, but it
doesn't make any difference to this discussion).
We will discuss coordinates further in Section \ref{ss:coordinates}.

From the same correlation function we can also define the cross-spectrum,
\begin{equation}
 P_\times(z,\Dtheta,k_\parallel) \equiv
  \int d \Delta v ~ e^{i \Delta v k_\parallel} 
  ~ \xi_{3D}(z,\Dtheta,\Delta v) =
  \int \frac{d\vk_\perp}{(2\pi)^2} ~ e^{i \vDtheta \vk_\perp} 
  ~ P_{3D}(z,k_\perp,k_\parallel) ~,
 \label{eq:PX}
\end{equation}
that quantifies the correlation of lines of sight modes as a function of 
angular separation.

We can now rewrite the elements of the covariance matrix in equation 
\ref{eq:covar} as a function of the cross-spectrum:
\begin{equation}
 C_{ij} = m_i ~ m_j \int \frac{d k_\parallel}{2\pi} 
    ~ e^{-i k_\parallel \Delta v_{ij}} 
    ~ P_\times(z_{ij},\Dtheta_{ij},k_\parallel) 
    ~ \tilde W_i(k_\parallel) ~ \tilde W_j(k_\parallel)
    + N_{ij} ~,
\end{equation}
where $\tilde W_i(k_\parallel)$ is the Fourier transform of the convolution 
kernel for pixel $i$, i.e., 
\begin{equation}
 \tilde W_i(k_\parallel) = e^{-k^2_\parallel G^2_i / 2} 
    ~ \frac{ \sin{ \left( k_\parallel T_i/2\ \right)} }{k_\parallel T_i/2} ~,
\end{equation}
where $G_i$ is the Gaussian spectral resolution and $T_i$ is the pixel width,
both in velocity units.\footnote{Physically, the resolution should be modeled as a continuous
convolution of the continuous field, only subsequently discretely sampled onto
pixels, but currently spectra are usually 
presented with a single resolution value per pixel, which is generally 
slowly varying from pixel to pixel, so we will not make
this distinction here}

\subsection{Parameterization}

We parameterize the power spectrum with a grid of band power parameters:
\begin{equation}
 P_{3D}(z,k_\perp,k_\parallel) =  P_{3D}^{\rm fid}(z,k_\perp,k_\parallel) 
      + \sum_\alpha ~ w^{3D}_\alpha(z,k_\perp,k_\parallel) ~ p^{3D}_\alpha ~, 
 \label{eq:P3D_param}
\end{equation}
where $P_{3D}^{\rm fid}(z,k_\perp,k_\parallel)$ is the fiducial power spectrum 
(around which the likelihood is expanded as explained in Section 
\ref{sec:Like}) and $w^{3D}_\alpha(z,k_\perp,k_\parallel)$ is a 
trilinear interpolation kernel.
Note that the fiducial power spectrum is evaluated at the full resolution,
independent of individual band-power bins. This decreases the impact
of features smaller than the bin size, which can still be represented 
perfectly if they are consistent with the fiducial model. 

We also consider an alternative parameterization in which we introduce a 
constant at each redshift that allows us to define a 3D magnitude 
of the wavevector,
\be
\tk\equiv \sqrt{\kpar^2+f^{-2}\left(z\right) \kperp^2}
\label{eq:tildek}
\ee
and cosine of the angle between the wavevector and the line of sight,
\be
\tmu \equiv \frac{\kpar}{\tk}~. 
\label{eq:tildemu}
\ee
$f\left(z\right)$ is necessary of course because observable $\kpar$ and 
$\kperp$ 
do not have the same units. 
We can then parameterize $\PtD$ by simple interpolation over $\tk$
and $\tmu$ -- we will call this $\PtDp$
\begin{equation}
\PtDp(z,\tk,\tmu) =  \PtDp^{\rm fid}(z,\tk,\tmu) 
      + \sum_\alpha ~ w^{3D^\prime}_\alpha(z,\tk,\tmu) ~ 
p^{3D^\prime}_\alpha ~, 
 \label{eq:P3Dp_param}
\end{equation}
We will discuss the relation between these parameterizations further below -- 
the advantage 
of $\tk,~\tmu$ parameterization is that if we choose $f(z)$ to be the 
Alcock-Paczynski factor \cite{1979Natur.281..358A} at each redshift for 
something close to the true model, then $\tmu$ will
be the cosine of the comoving angle corresponding to observable 
$(\kpar,~\kperp)$. 
The BAO feature will be approximately independent of $\tmu$
which should allow relatively coarse interpolation in $\tmu$, and the linear 
redshift space boost of power  will be approximately independent of $\tk$, 
producing aesthetically pleasing plots.

As discussed in Section \ref{sec:numeric}, we will start our analysis by 
estimating the likelihood as a function of cross-spectrum parameters, 
defined as:
\begin{equation}
 P_\times(z,\Dtheta,k_\parallel) =  P_\times^{\rm
   fid}(z,\Dtheta,k_\parallel) + \sum_\alpha ~ \Theta_\alpha(\Dtheta) 
    ~ w^\times_\alpha(z, k_\parallel) ~ p^\times_\alpha ~, 
 \label{eq:Px_param}
\end{equation}
where $w^\times_\alpha(z, k_\parallel)$ is a bilinear interpolation kernel 
and $\Theta_\alpha(\Dtheta)$ is a top-hat kernel in the transverse direction.
We chose a nearest-grid-point (NGP) interpolation in the transverse direction 
because, in combination with the numerical approximation described in 
\ref{ss:block}, will reduce considerably the number of elements of the Fisher 
matrix that we need to compute. 
Using NGP in $w^\times_\alpha(z, k_\parallel)$, on the other hand, would 
not reduce it further. 
This will be discussed again in \ref{ss:block}.

We will discuss after explaining our detailed numerical algorithm why we do 
not from the start define bands in approximate comoving
coordinates using a fiducial cosmology. 
Suffice it to say at this point that
this is not necessary as long as we plan to make measurements in sufficiently
narrow redshift bins. As long as we use a good interpolation scheme, i.e., 
one that converges to an exact representation of the function in the limit
of fine point-spacing, using different coordinates to represent the same 
function can only improve the efficiency of the representation, i.e., allow 
us to represent the function accurately with fewer interpolation points. 
We will always test explicitly that results of interest have converged 
with respect to band spacing. 

The covariance matrix is given by 
$\vC = \vC^{\rm fid}+ \vC_{,\alpha} ~ p^\times_\alpha + \vN $, 
with the derivative with respect to one of these band power parameters being:
\begin{equation}
 \vC_{ij,\alpha} 
    = m_i ~ m_j ~ \Theta_\alpha(\Dtheta_{ij}) 
      \int ~ \frac{d k_\parallel}{2\pi} 
      ~ e^{i k_\parallel \Delta v_{ij}} 
      ~ \tilde W_i(k_\parallel) ~ \tilde W_j(k_\parallel) 
      ~ w_\alpha(z_{ij}, k_\parallel) ~,
 \label{eq:dCdp}
\end{equation}
and we have all the pieces required to compute the derivatives of the 
(log-) likelihood described in section \ref{sec:Like}.

Note that each matrix $\vC_{,\alpha}$, corresponding to a cross-spectrum
parameter $p^\times_\alpha$ with angular separation $\Dtheta_\alpha$, will 
have a very sparse structure, since only the sub-matrices corresponding to 
pairs of spectra separated by $\Dtheta_{ij} \in \Dtheta_\alpha$ will be 
non-zero.
This would not be the case if we wanted to directly estimate the
three-dimensional power, where a given pair of pixels would contribute to
all band power parameters.

\subsubsection{Inverting functional parameterizations}

We have introduced several parameterization of continuous functions of the
form $P(k) = P^{\rm fid}(k)+\sum_\alpha w_\alpha(k) p_\alpha$, where 
we use $k$ here to represent all continuous variables. 
As usual, we describe the likelihood to be a function of data given a set
of parameters $\vp$, $L(\vd|\vp)$. 
Theories generally predict $P(k)$, so to use this kind of 
likelihood function we need to know how to compute $\vp[P(k)]$.
We intend to work mostly close to the limit of  
high resolution parameterization, i.e., $k$ spacing between $p_\alpha$ points
smaller than any structure in $P(k)$, where it should be fairly obvious that
we can simply take $p_\alpha=P(k_\alpha)-P^{\rm fid}(k_\alpha)$. 
However, to approach this limit 
robustly it is useful to think a little more carefully about this problem.
A formal solution is 
$p_\alpha=\int dk~ w_\alpha^{-1}(k) \left[P(k)-P^{\rm fid}(k)\right]$, viewing 
$w_\alpha(k)$ as an $\infty\times N_p$ matrix, however, this matrix is not
generally invertible. A natural thing to do is use the pseudo-inverse of
$w_\alpha(k)$, equivalent to finding $\vp$ to give the least 
square deviation from the target $P(k)$. Assuming equal weight for all
$k$, i.e., using uniform discretization to make $w_\alpha(k)$ into a finite
dimensional matrix, 
the solution is:
\be
p_\alpha = \sum_\beta \left(I^{-1}\right)_{\alpha\beta} \int dk~ 
\left[P(k)-P^{\rm fid}(k)\right] w_\beta(k) 
\label{eq:inverting}
\ee
with $I_{\alpha\beta}=\int dk~ w_\alpha(k)w_\beta(k)$. 
We see that $I^{-1}w$ is the response of the parameter to a delta function 
impulse in $P(k)$ away from $P^{\rm fid}(k)$.  In the simple case of 
top-hat bands which do not overlap, this reduces to simply averaging 
$P(k)$ over the band, but for other cases like linear interpolation it is not
quite so simple. E.g., for two parameters at $k=0$ and $k=1$ determining the
$0<k<1$ interval by linear interpolation, the effective weight on $P(k)$ for
the first ($k=0$) parameter is $4-6 k$, i.e., it peaks at $k=0$ and goes 
linearly through zero at $k=2/3$ to become negative at higher $k$. Intuitively,
if the power is high at $k \sim 1$, the 2nd parameter will be larger, requiring
a lower value of the first parameter to describe the intermediate
range. When more bins are considered, $I$ is a band matrix with just
one additional diagonal, but $I^{-1}$ is not, coupling elements that
are more than one bin apart. 
In less trivial setups we can always evaluate the formula numerically, but we 
see also that in the
limit of narrow $w_\alpha(k)$, relative to the structure in $P(k)$, we are
justified using $p_\alpha=P(k_\alpha)-P^{\rm fid}(k_\alpha)$. 

Another approach
to this problem is to compute the expected response of estimated parameter
values
to changes in the true model in much narrower bands 
\cite{1999PhRvD..60j3516K,2012ApJ...761...13S}, which
amounts to going back to the data to re-compute the Fisher matrix with one
$\partial\vS/\partial\vp$ 
replaced with derivatives with respect to the much finer bands. 
While it is possible that this approach could save some computation over 
simply making the original
measurement bands themselves narrower, in this paper our plan is to always 
make sure that our end results have converged with respect to the width of the
measurement bins, i.e., we will make our measurement bands narrow enough that
the detailed structure within them is not important. 
A natural question at this point is: what is the relationship between this 
data-derived method and the mathematical pseudo-inverse above? 
The key is that
the discretization required for the pseudo-inverse, assumed to be uniform in
$k$ in Eq. (\ref{eq:inverting}), is ambiguous as a matter of pure math.  
Another way to
look at the pseudo-inverse is that if $P$ was measured using many finely and 
uniformly spaced
points with equal errorbars, then Eq. (\ref{eq:inverting}) 
would give the best fit for the coarsely
interpolated parameterization. The data-derived method determines a more 
correctly representative, in general non-uniform, weighting within the bins. 

\subsection{Connection to the traditional 1D power spectrum}

In the limit of zero angular separation, the cross-spectrum becomes the
traditional 1D power spectrum:
\begin{equation}
 P_{1D}(z,k_\parallel) \equiv P_\times(z,\Dtheta=0,k_\parallel)
  = \int \frac{d\vk_\perp}{(2\pi)^2} ~ P_{3D}(z,k_\perp,k_\parallel) ~.
\end{equation}

It is very common in 1D analyses to ignore the correlation between the
fluctuations in neighboring spectra, resulting in a block-diagonal
covariance matrix and in a likelihood that is factorizable.  The
derivatives of the log-likelihood are then additive, and we recover
the method to estimate the 1D power spectrum presented in
\cite{2006ApJS..163...80M} and later used in
\cite{2013A&A...559A..85P}.\footnote{Note, however, that the redshift
  interpolation of the power is better defined in our analysis. For
  instance, \cite{2013A&A...559A..85P} split each spectrum into two or
  three segments, and assumed that each segment contributed to a
  single redshift bin.}

\section{Numerical implementation}
\label{sec:numeric}

A naive implementation of the 
algorithm implied by sections \ref{sec:Like} and \ref{sec:Lya} is 
computationally
intractable, since 
it involves thousands of linear algebra operations with products of extremely 
large matrices, of size set by the number of pixels in the 
survey ($\approx 10^8$ for relevant surveys).

In this section we will discuss some approximations and numerical
implementations that make the problem tractable, and that allow us to
analyze the BOSS data set typically using a few thousand CPU hours
depending on the number of bins used.

We will start in \ref{ss:PX} by choosing a parameterization of the likelihood
in which a quasar pair contributes only to a small fraction of parameters.
In \ref{ss:block} we present a particular fiducial clustering model that 
allows us to have a block-diagonal covariance matrix, reducing drastically 
the size of the linear algebra operations required. 
In \ref{ss:eigen} we show that the information in the response matrices can
be described by a very small number of eigenmodes.
Finally, in \ref{ss:FFT} we further speed up the algorithm by internally
using Fast Fourier Transforms (FFT) to compute some of the matrix 
multiplications.

\subsection{Cross-spectrum}
\label{ss:PX}

We aim to estimate first the cross-spectrum, because using it each pair of
quasars contributes only to a single parameterized band in the transverse 
coordinate 
(angular separation), instead of contributing to all parameters if we went
straight to a 3D power spectrum. 
Estimating $P_\times$ from the raw data 
is the computationally demanding step. After that, the data set is massively 
compressed and we have a lot of options to transform to other 
parameterizations, which we will discuss below. To be clear: it is the 
sparsity of the quasars on the sky in real data sets, relative to the 
scales of interest, that motivates 
approaching them pair-wise, and therefore using the cross-spectrum. 
This is analogous to the efficiency of 
estimating a 3D correlation function pair-wise for a sparse set of point 
objects (which could similarly be converted to a power spectrum by the methods
below). 
If we had
much denser sampling (or wanted to efficiently push the measurement to much
larger scales) we would be motivated to use a method that, e.g., compressed
spectra in an angular pixel into some aggregate measure of the flux in that 
pixel, transformed in the transverse directions, and then measured the 3D power 
spectrum directly. To cover both regimes
we could imagine something akin to the P$^3$M approach to N-body simulations
(e.g., \cite{2013MNRAS.436..540H,2007arXiv0711.4655S,2003NewA....8..581P} -- 
generally, any of the ways similar problems have been solved 
for N-body simulations could be useful).

\subsection{Zero cross-correlation fiducial model gives block-diagonal
$\vC_0^{-1}$ \label{ss:block} }

In order to measure 3D correlations we need to consider the full
likelihood, and we need to choose a fiducial power 
$P_\times^{\rm fid}(z,\Dtheta,k_\parallel)$ 
in order to specify the global weighting matrix $\vC_0$.  The ideal
choice would be to use our global best guess of the \emph{true} power
spectrum, a $\Lambda$CDM model with linear bias and redshift space
distortions and appropriate non-linear correction on smaller scales
\cite{2003ApJ...585...34M,2015JCAP...12..017A, 2017A&A...603A..12B}. 
If we choose a different model, our Taylor series generally will not be as 
accurate near the true model as discussed in Section \ref{sec:Like}.

%In this work we take the fiducial power spectrum to be
In this work we take the fiducial cross-spectrum to be
\begin{equation}
   P_\times^{\rm fid}(z,\Dtheta,k_\parallel)= P^{\rm
     fid}_{1D}(z,k_\parallel) ~ \delta^D(\Dtheta),
\end{equation}
i.e. it is vanishing everywhere except at zero angular separation. 
This makes the weighting matrix block-diagonal giving a huge speed up.
Because variance of the field is dominated by very small
scale modes, or equivalently, because the $P_\times$ is very rapidly
falling with $\Dtheta$ 
(as shown in Figure \ref{fig:PX} below), this is a reasonably good 
approximation to the truth. 
This is further supported by the study presented in Section \ref{sec:P3D} 
below, and in particular in Figure \ref{fig:Pzlkp_chi2_N}, where we show that 
any possible underestimation of the measured uncertainties caused by this 
approximation is smaller than 6-8\%.

We use subscripts $I$ and $J$ to identify blocks corresponding to a 
particular quasar. 
We can now describe the (log-) likelihood derivatives as sum over blocks:
\begin{equation}
 \sL_{,\alpha}=\frac{1}{2} \sum_I \sum_J \vy_I^t ~ \vC_{IJ,\alpha} ~ \vy_J 
    - \frac{1}{2} \sum_I {\rm Tr}\left[\vC_{0~I}^{-1} ~ \vC_{II,\alpha}\right]~,
 \label{eq:dL1_PX}
\end{equation}
and
\begin{equation}
 \sL_{,\alpha\beta} \approx - F_{\alpha\beta} = - \frac{1}{2} \sum_I \sum_J 
    {\rm Tr} \left[\vC_{0~I}^{-1} ~ \vC_{IJ,\alpha} ~
    \vC_{0~J}^{-1} ~ \vC_{JI,\beta}\right]~,
 \label{eq:dL2_PX}
\end{equation}
where we have also defined $\vy = \vC^{-1}_0 \left(\vd-\vm\right)$ and 
$\vC_{IJ,\alpha}$ is the sub-matrix of $\vC_{,\alpha}$ corresponding to 
the cross-correlation of pixels in quasars $I$ and $J$.
From equation \ref{eq:dL1_PX} above it is clear that the second term of the 
first derivative, the one with a trace, will only be non-zero for derivatives
with respect to $P_{1D}$ parameters. With our top-hat parameterization in the
$\Delta \theta$ direction, the Fisher matrix is non-zero only for
parameters corresponding to the same value of $\Delta \theta$.

\subsection{Eigenmode decomposition of $\vC_{,\alpha}$ matrices}
\label{ss:eigen}

We started with a problem that was computationally impossible, and thanks to 
the choice of parameters ($\vp^\times$ instead of $\vp^{3D}$) and the choice 
of weighting matrices (block diagonal $\vC_0$) we have now a problem that is 
just extremely computationally intensive. 

Let us consider a BOSS like survey, with an area of 10 000 square degrees 
and a density of lines of sight of roughly 15 per square degree. 
If we are interested in measuring 3D correlation up to 3 degrees 
(roughly $200 \hMpc$), we will need to correlate each spectrum with an 
average of 400 neighbors, for a total of roughly 30 million pairs of spectra
(not counting the same pair twice).
If we are measuring a total of 1000 cross-power spectrum parameters 
(as discussed below we typically use several thousand parameters), the Fisher
matrix (or second derivative) will have a million elements $F_{\alpha\beta}$,
and even if we consider only the most relevant covariances we still need to 
compute several tens of thousand elements. 
As shown in equation \ref{eq:dL2_PX}, for each element $F_{\alpha\beta}$ we 
need to compute the trace of 30 million products of four matrices, of size 
equal to the number of pixels in a spectrum, roughly 500.  

The algorithm described above would work on a small data set, with limited 
number of parameters, but it quickly becomes too slow when applied to any 
relevant survey. 
In this section we will introduce numerical implementations that will speed up
dramatically the algorithm without adding further approximations.

To motivate the math below, let us consider first the simplified example of 
a measurement on a very large periodic box, with parallel lines of 
sight and a static universe, and ignoring pixelization or resolution effects. 
In this case we could define the following modes for each line of sight:
\begin{equation}
 \tilde \delta(\theta,k_\parallel) = \int_{-\infty}^{\infty} ~ dv ~ 
        e^{i k_\parallel v} ~ \delta(\theta,v) ~,
 \label{eq:mode1D}
\end{equation}
and it is easy to show that these modes would be uncorrelated for different 
values of $k_\parallel$, and that their correlations would be given by the 
cross-power spectrum:
\begin{equation}
 \left< \tilde \delta(\theta,k_\parallel) 
        ~ \tilde \delta(\theta^\prime,k_\parallel^\prime) \right> 
  = (2 \pi) ~ \delta^D(k_\parallel+k_\parallel^\prime) 
        ~ P_\times(\Dtheta, k_\parallel) ~,
 \label{eq:idealPX}
\end{equation}
where $\delta^D(\kpar)$ is the Dirac delta function.
In this space, both the covariance matrix $\vC_{0~I}$ and its derivatives
$\vC_{IJ,\alpha}$ would be diagonal. 
Redshift evolution and the finite length of the lines of sight in a realistic
survey complicate this picture, but as we show below most of the information
in the matrices $\vC_{IJ,\alpha}$ is still captured in a small number of 
modes.

So we would like to project each spectrum onto some kind of Fourier-like
modes, but while the parameter response matrices $\vS_{,\alpha}$
are localized in redshift, Fourier modes are not. 
For this reason we could hope to do better by using Fourier transforms over a 
set of ad hoc redshift windows.
What we do will be qualitatively similar to this, but with the modes and
envelopes more carefully derived.  
Our first thought to derive modes might be to use Karhunen-Loeve (KL) 
eigenmodes
\cite{1997ApJ...480...22T}, which are generally eigenmodes of  
$\vL^{-1} \vC_{,\alpha}\vL^{-t}$, where $\vC=\vL \vL^t$ is a Cholesky 
decomposition.  
As we have set things up, each quasar pair
would have its own set of 
$\vL^{-1}\vC_{IJ,\alpha}\vL^{-t}$ matrices, covering a different set 
of wavelengths, and with different pixelization and resolution. 
We would need to compute the eigenmode decomposition of all of them,
which is easier than computing KL modes for the fully coupled data set, but 
would still make the algorithm intractable. 

For this reason, we will first extend a little bit our data model. 
Let us define a common pixel grid covering the whole wavelength range of 
the survey.\footnote{In particular we use a grid of 2450 cells, of same width as in BOSS 
coadded spectra $dv=69.03 \kms$, covering the redshift range $1.74<z<3.81$.}
We can map all our spectra into that grid, giving effectively infinite noise 
to cells that are not covered by a given spectrum (or masked by the pipeline).
If we discretize the integral over wavelength in equation \ref{eq:data_vector}
we can rewrite the data model as:
\begin{equation}
 \vd = \vR \vdelta + \vm + \vep ~,
\end{equation}
with $R_{ij} = m_i ~ W_i(\lambda_j)$, and the covariance matrix is:
\begin{equation}
 \vC = \left<\left(\vd-\vm\right) \left(\vd-\vm\right)^t\right> 
    = \vR \vS \vR^t +\vN
\end{equation}
where $\vS$ is now the signal covariance without pixel or resolution effects.\footnote{To prevent numerical artifacts when evaluating Equation \ref{eq:dSdp}
we compute $\vS$ with a small Gaussian smoothing of $G_0=20 \kms$, and 
correct for it by using $\sqrt{G_i^2 - G_0^2}$ as the pixel resolution when 
computing the resolution matrices $\vR$. We have checked that the results do 
not depend on the exact value of $G_0$ used.}

The derivatives of the covariance matrix for a pair of spectra are now:
\begin{equation}
 \vC_{IJ,\alpha} = \vR_I \vS_{IJ,\alpha} \vR_J^t ~,
\end{equation}
and similar to equation \ref{eq:dCdp} we can write them as:
\begin{equation}
 S_{IJij,\alpha} = \Theta_\alpha(\Dtheta_{IJ}) 
        \int \frac{d k_\parallel}{2\pi} ~ e^{-i k_\parallel \Delta v_{ij}}  
        ~ w_\alpha(z_{ij}, k_\parallel) 
      \equiv \Theta_\alpha(\Dtheta_{IJ}) ~ \tilde S_{ij,\alpha} ~.
 \label{eq:dSdp}
\end{equation}

\begin{figure}[H]
 \begin{center}
  \includegraphics[scale=0.5]{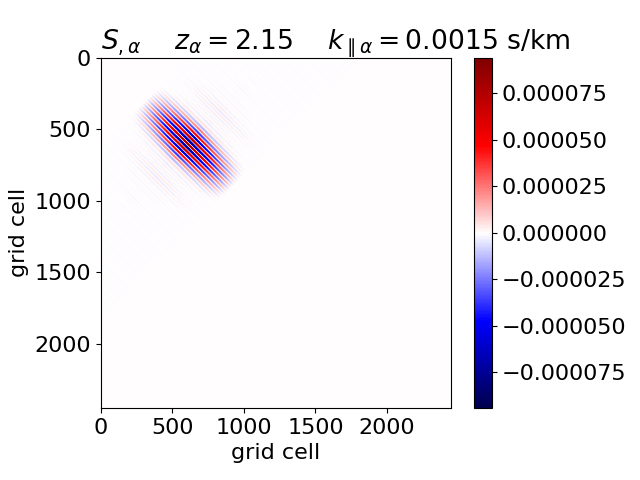}
  \includegraphics[scale=0.5]{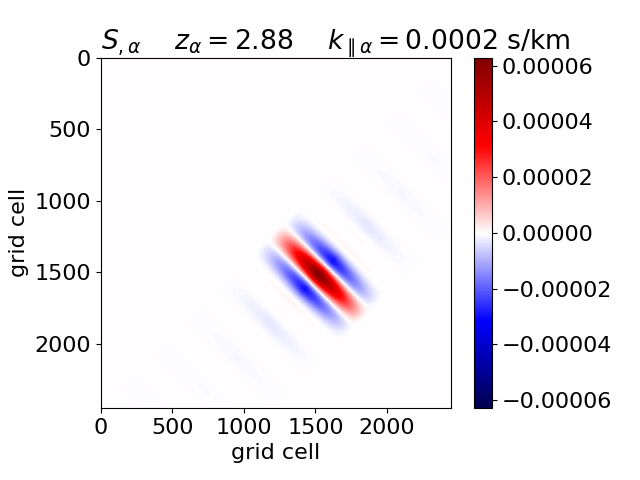}
 \end{center}
 \caption{
  $\tilde \vS_{,\alpha}$ matrices for two different parameters. Left: 
  eighth k-bin ($k_\parallel=0.0015 \ikms$) of the second z-bin ($z=2.15$);
  Right: first non-zero k-bin ($k_\parallel=0.0002 \ikms$) of the fifth z-bin 
  ($z=2.88$). 
  The redshift bin (z-bin) of the parameter sets the diagonal band that is
  non-zero, and the wavenumber bin (k-bin) sets the oscillation frequency.
 }
 \label{fig:dSdp}
\end{figure}

Equation \ref{eq:dSdp} implies that: i) the derivative matrix $\vS_{IJ,\alpha}$
will be zero unless $\Dtheta_{IJ} \in \Dtheta_\alpha$ and ii) if non-zero, the 
matrix $\tilde \vS_{,\alpha}$ will only depend on ($z_\alpha$,$k_\alpha$) and 
will be the same for all quasar pairs.  
Therefore, if we choose to project onto eigenmodes of $\tilde \vS_{,\alpha}$, 
we will only need to decompose as many matrices 
$\tilde \vS_{,\alpha}$ as parameters in our analysis.
In Figure \ref{fig:dSdp} we show a couple of $\tilde \vS_{,\alpha}$
matrices that might appear in a typical analysis.

Now we simply {\it exactly diagonalize} $\tilde \vS_{,\alpha}$, i.e., find
$\tilde \vS_{,\alpha} = \vT_\alpha \vD_\alpha \vT^t_\alpha$ where $\vD_\alpha$ 
is diagonal.  This is motivated by the fact that in the limit of
infinitely narrow $k$-bins and infinitely large redshift bin, these
signal matrices would have exactly two non-zero
eigenvalues.\footnote{To see this, note that a single Fourier mode of any phase will 
 have correlation function $\xi(\Delta v) \propto \left<\sin(\phi) 
 \sin(\phi+k \Delta v)\right>\propto \cos(k \Delta v)$ and that
 $C_{ab}\propto\cos(k (a-b) ~ dv)
  = \cos(k a ~ dv) \cos(k b ~ dv) + \sin(k a ~ dv) \sin(k b ~ dv)$, 
where $dv$ is the cell width of the grid. 
In this case the pixel covariance can be written as
$C(r)\propto[\cos(kv)]^T[\cos(kv)] + [\sin(kv)]^T[\sin(kv)]$.
}
A matrix with finite redshift and $k$-bins is
thus expected to have a few dominant eigenvectors.

\begin{figure}[H]
 \begin{center}
  \includegraphics[scale=0.5]{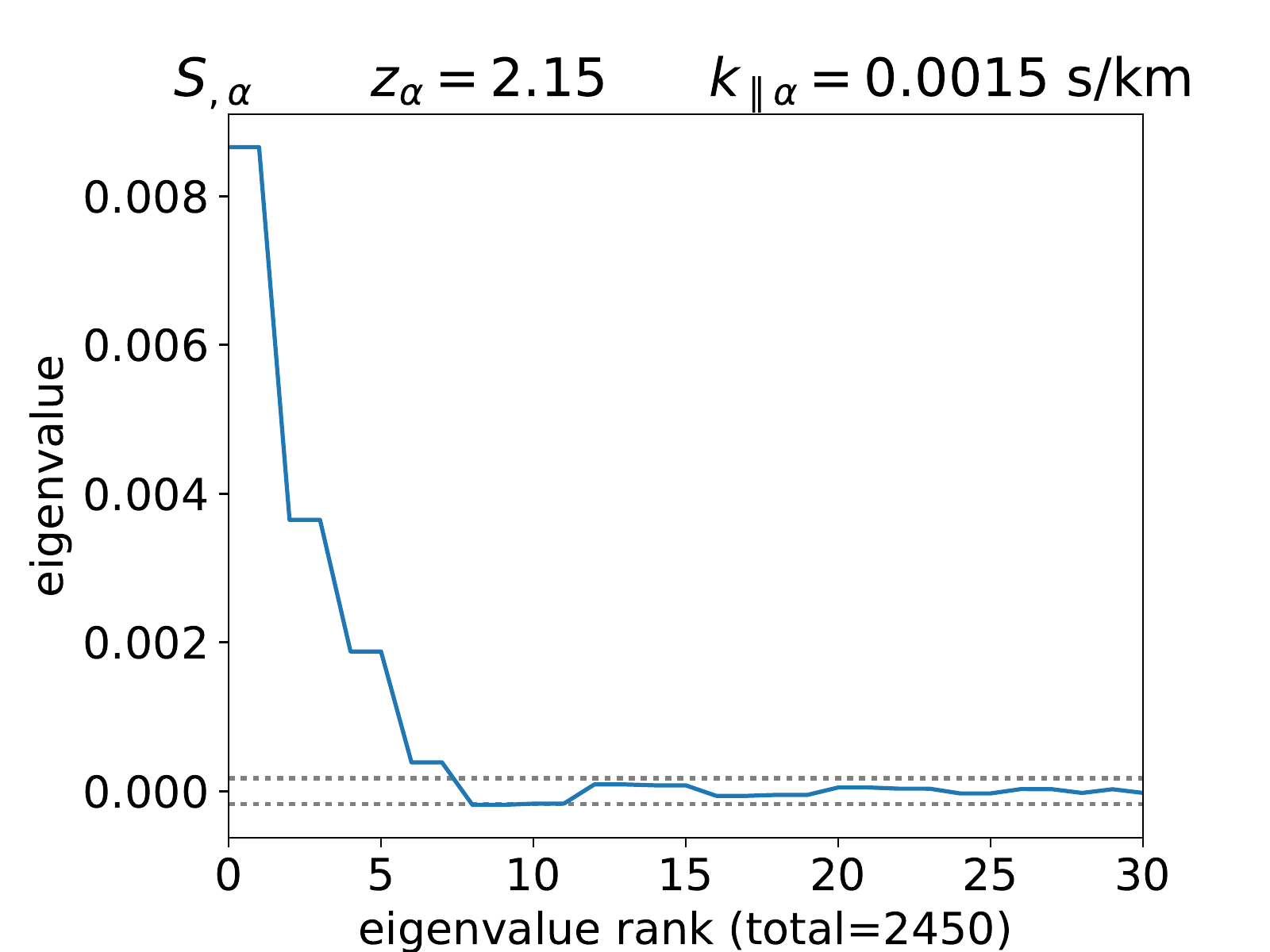}
  \includegraphics[scale=0.5]{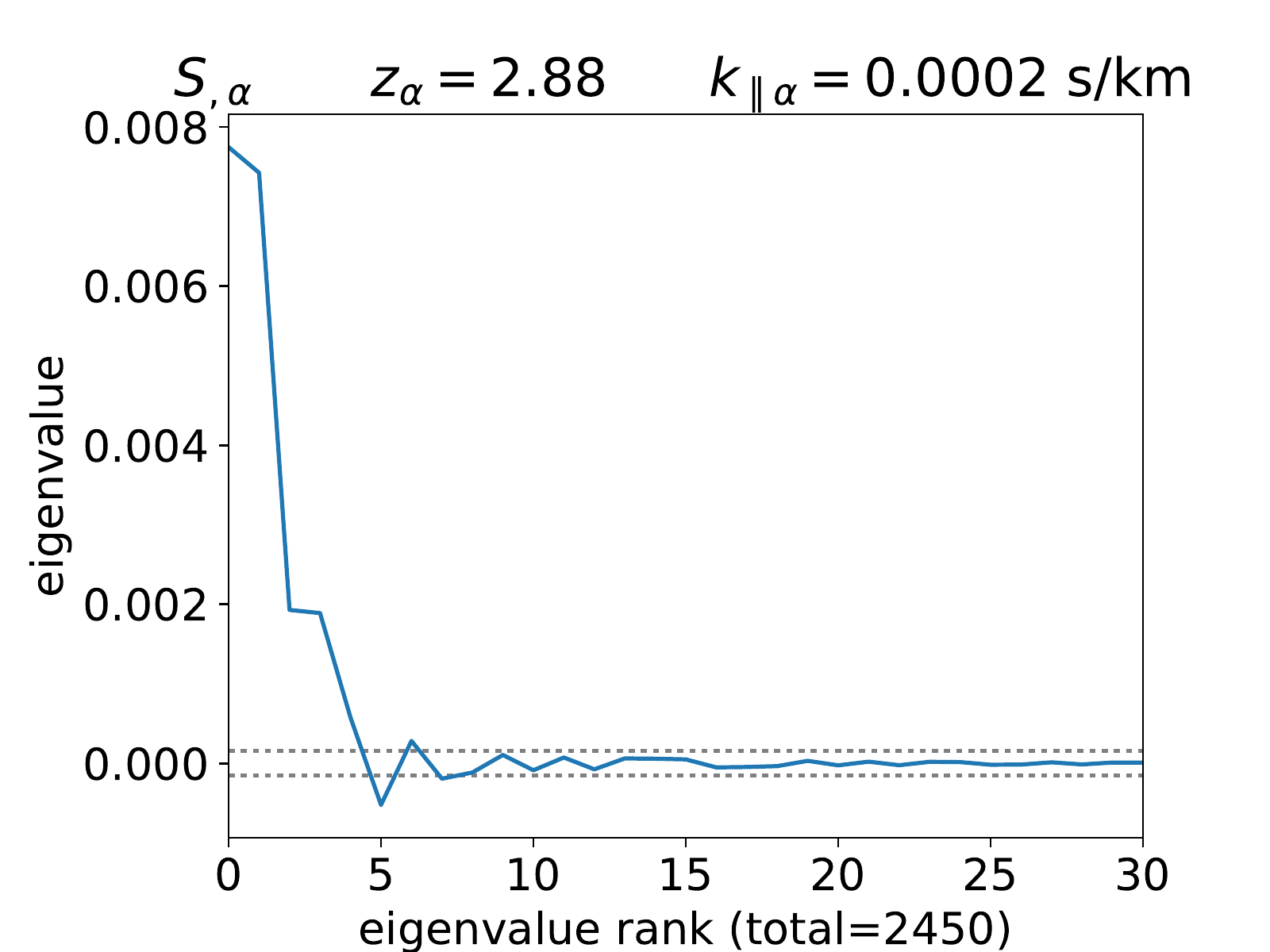}
 \end{center}
 \caption{Eigenvalues for the two $\tilde \vS_{,\alpha}$ matrices shown in 
  Figure \ref{fig:dSdp}, sorted by absolute value (the total number of 
  eigenvalues is 2450). 
  The dotted lines show the $\pm 2\%$ of the maximum eigenvalue, and we ignore
  eigenmodes with an eigenvalue smaller than that (in absolute value).
Note that negative eigenvalues are legitimate, reflecting the fact
that $\tilde \vS_{,\alpha}$ is not itself a legitimate (positive definite)
covariance matrix. 
 }
 \label{fig:eigval_dSdp}
\end{figure}

In Figure \ref{fig:eigval_dSdp} we present the distribution of eigenvalues for 
the $\tilde \vS_{,\alpha}$ matrices presented in figure \ref{fig:dSdp}, 
sorted by absolute value. 
In both cases the number of relevant eigenvalues is very small (6 or 7 
out of 2450).
The dotted lines show the $\pm e_{\rm min}=0.02$ of the maximum eigenvalue, 
and we generally 
neglect those eigenmode with an eigenvalue smaller (in absolute value) 
than that.
This means that not only the $\vD_\alpha$ matrices presented above are diagonal,
but only a handful of elements are considered non-zero. 

\begin{figure}[H]
 \begin{center}
  \includegraphics[scale=0.5]{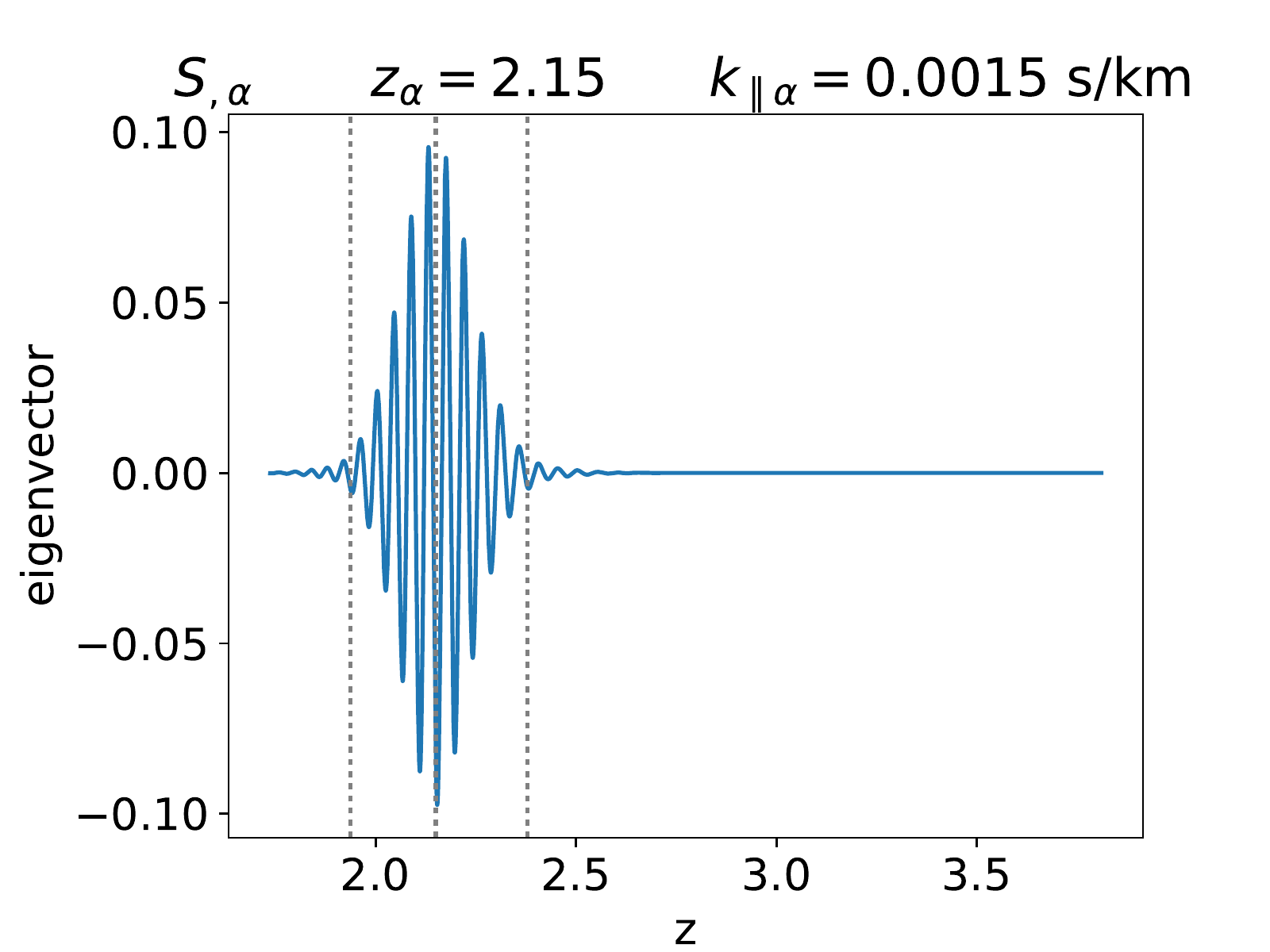}
  \includegraphics[scale=0.5]{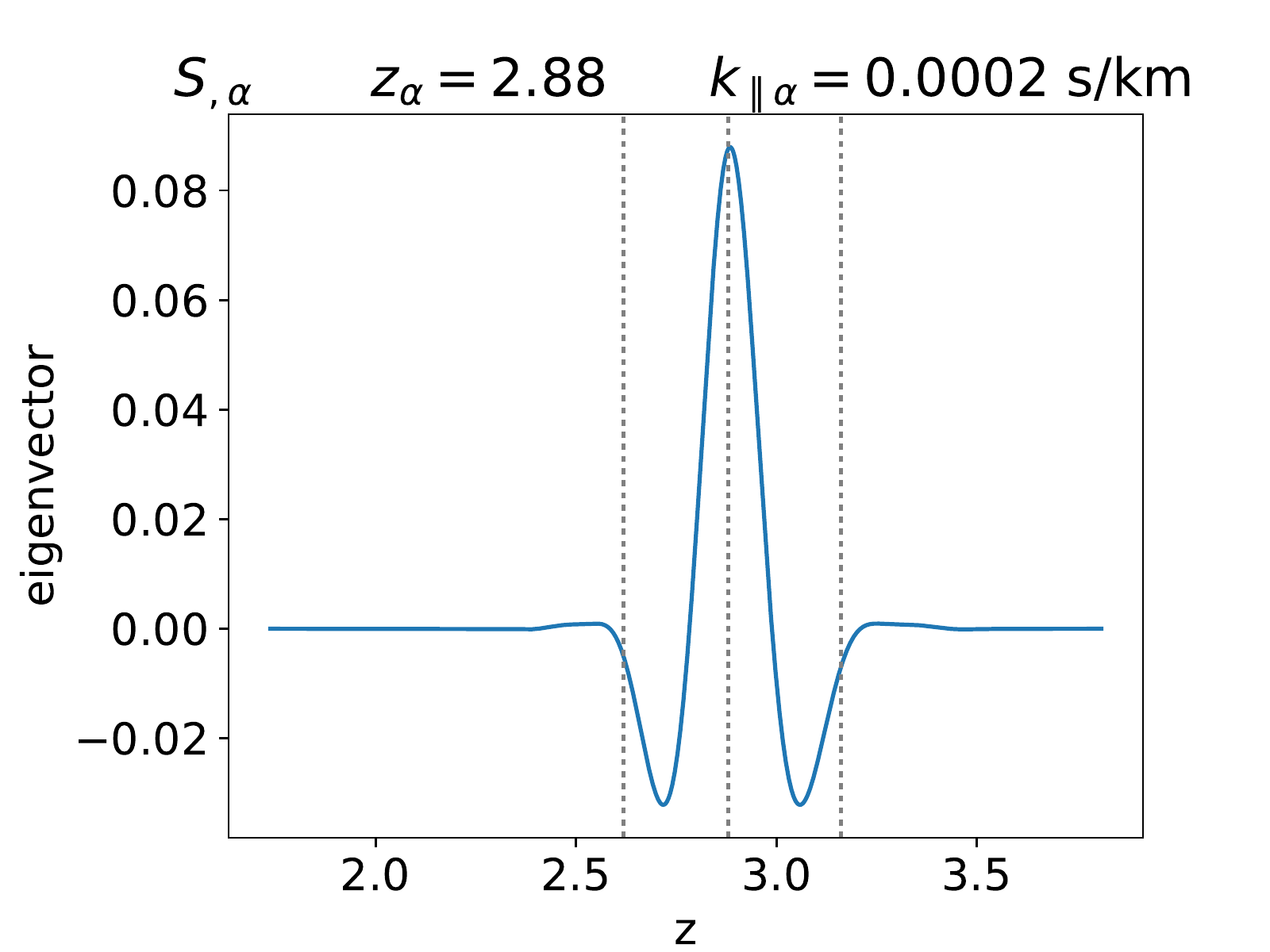}
 \end{center}
 \caption{First eigenvector for the $\tilde \vS_{,\alpha}$ matrices in 
  figure \ref{fig:dSdp}.
  The vertical dotted lines show the center and the edges of the redshift bin 
  $z_\alpha$ for the parameter.
  The wavenumber $k_{\parallel \alpha}$ of the parameter in the left panel 
  is $7.5$ times larger than the one in the right panel.
 }
 \label{fig:PLA_T}
\end{figure}

In Figure \ref{fig:PLA_T} we show the first eigenvector for the same matrices
$\tilde \vS_{,\alpha}$. 
The left panel corresponds to a parameter with wavenumber $7.5$ times larger 
than the wavenumber in the right panel.
They are both localized in a region of the grid corresponding to the redshift 
of the parameter in the derivative, as shown by the vertical dotted lines.

We will compute the eigendecomposition only once per parameter, and use the 
decomposition for all quasars pairs. 
With this diagonalization we can now look again at the derivatives. 
The first term in the first derivatives, i.e., the one involving the data 
vector, will now be:
\begin{align}
 \vy^t_I ~ \vC_{IJ,\alpha} ~ \vy_J 
  & = \vy^t_I ~ \vR_I ~ \tilde \vS_{,\alpha} ~ \vR_J^t ~ \vy_J  \nonumber \\
  & = \vy^t_I ~ \vR_I ~ \vT_\alpha \vD_\alpha \vT^t_\alpha 
      ~ \vR^t_J ~ \vy_J \nonumber \\
  & = \vy^t_{I\alpha} ~ \vD_\alpha ~ \vy_{J\alpha} ~,
\end{align}
where $\tilde \vS_{,\alpha}$ is the same for all quasars pairs that contribute 
to this particular angular separation bin, 
$\vy_I = \vC_{0~I}^{-1} (\vd_I - \vm_I)$ is now formally defined over all
wavelengths but it is approximately limited to the range actually covered by 
the spectrum, and we have defined 
$\vy_{I\alpha} = \vT^t_\alpha ~ \vR_I^t ~ \vy_I$. For each parameter,
$\vy_{I\alpha}$ is akin to the collection of Fourier modes in the $k$ range
corresponding to the parameter, for a transform windowed to the $z$ range 
covered by the parameter.

Similarly we can expand the second term of the first derivative:
\begin{align}
 {\rm Tr} \left[\vC_{0~I}^{-1} ~ \vC_{II,\alpha}\right]       
  & = {\rm Tr} \left[\vC_{0~I}^{-1} 
      ~ \vR_I ~ \tilde \vS_{,\alpha} ~ \vR_I^t \right]          \nonumber \\
  & = {\rm Tr} \left[\vC_{R~I}^{-1} ~ 
      ~ \vT_\alpha \vD_\alpha \vT^t_\alpha \right] \\
  & = {\rm Tr} \left[\vM_{I\alpha\alpha} \vD_\alpha \right]
\end{align}
and the second derivative (or Fisher matrix):
\begin{align}
 {\rm Tr} \left[\vC_{0~I}^{-1} ~ \vC_{IJ,\alpha} 
      ~ \vC_{0~J}^{-1} ~ \vC_{JI,\beta}\right]
  & = {\rm Tr} \left[\vC_{0~I}^{-1} ~ \vR_I ~ \tilde \vS_{,\alpha} ~ \vR_J^t
      ~ \vC_{0~J}^{-1} ~ \vR_J ~ \tilde \vS_{,\beta} ~ \vR_I^t \right] \nonumber \\
  & = {\rm Tr} \left[\vC_{R~I}^{-1} ~ \vT_\alpha \vD_\alpha \vT^t_\alpha
      ~ \vC_{R~J}^{-1} ~ \vT_\beta \vD_\beta \vT^t_\beta \right] \nonumber \\
  & = {\rm Tr} \left[\vM_{I\beta\alpha} ~ \vD_\alpha 
      ~ \vM_{J\alpha\beta} ~ \vD_\beta \right]  ~,
 \label{eq:F_eigen}
\end{align}
where we have defined $\vC_{R~I}^{-1} = \vR_I^t \vC_{0~I}^{-1} ~ \vR_I$ and 
$\vM_{I\alpha\beta} = \vT^t_\alpha ~ \vC_{R~I}^{-1} ~ \vT_\beta$.
We will have to compute the objects $\vy_{I\alpha}$ and $\vM_{I\alpha\beta}$
only once for each spectrum, and once we have them the cross-correlations will 
be very fast.

Note that since we only consider $N \sim 7$ eigenmodes (see Figure 
\ref{fig:eigval_dSdp}), the diagonal matrices $\vD_\alpha$ have only $N$ 
non-zero elements. 
This means that we only need to compute a very small block $N \times N$ of 
elements of the $\vM_{I\alpha\beta}$ matrices.

\subsection{FFT of eigenvectors}
\label{ss:FFT}

The eigenmode decomposition described above moved the computational challenge
from the spectra cross-correlations to the computation of \textit{per-spectrum}
objects: $\vy_{I\alpha}$ and $\vM_{I\alpha\beta}$. 
In particular the slowest part is to compute the rotated inverse covariances:
\begin{equation}
 \vM_{I\alpha\beta} = \vT^t_\alpha ~ \vC_{R~I}^{-1} ~ \vT_\beta ~,
\end{equation}
for which naively we need to compute a product of three matrices of size equal 
to the number of cells in our radial grid, 2450.
For each quasar $I$, naively we need to compute this product for each pair
of matrices ($\vT_\alpha$,$\vT_\beta$), specified by their values of 
($z$, $\kpar$).\footnote{Remember that in Equation \ref{eq:dSdp} we introduced 
$\tilde \vS_\alpha$ as matrices that do not depend on $\Dtheta_\alpha$.}
In the analysis presented in Section \ref{sec:PX}, this would imply over 
$10^4$ matrices per quasar. On the other hand, we know that in the idealized
high symmetry case this rotation amounts to 
FFTs along each row and column which can be done orders of magnitude faster.

Fortunately, matrices $\vM_{I\alpha\beta}$ are only used to compute the elements
of the Fisher matrix $F_{\alpha\beta}$, as shown in Equation \ref{eq:F_eigen}.
Parameters that have very different values of either $z$ or $\kpar$ will be
very weakly correlated, and therefore we decide to compute only a subset of 
the elements of the Fisher matrix.
In particular, we ignore the correlation of parameter pairs with either
$| \Delta \kpar | > F_k$ or 
$| \Delta \ln{(1+z)} | > F_z + 2\pi/(c ~ \bar \kpar)$, where $\bar \kpar$ is the
geometric mean and $c$ is the speed of light, with the exception of 
neighboring parameters that are always included.\footnote{The last term was added to allow correlations of the lowest values
of $\kpar$ over all redshift ranges.}
In our default analysis, this reduces the number of $\vM_{I\alpha\beta}$ 
matrices we need to compute by an order of magnitude. 

The eigenvectors shown in figure \ref{fig:PLA_T} have clear oscillatory 
features, what suggests that they would be very sparse in Fourier space. 
For this reason we introduce discrete Fourier transform and inverse Fourier
transform matrices pairs $\vF ~ \vF^{-1}$ in the computation of the per-quasar
objects:
\begin{align}
 \vM_{I\alpha\beta} & = \vT^t_\alpha ~ \vC_{R~I}^{-1} ~ \vT_\beta  \nonumber \\
  & = \vT^t_\alpha ~ \vF^t ~ \vF^{-t} ~ \vC_{R~I}^{-1} 
      ~ \vF^{-1} ~ \vF ~ \vT_\beta                                 \nonumber \\
  & = \tilde \vT^t_\alpha ~ \tilde \vC_{R~I}^{-1} ~ \tilde \vT_\beta ~,
 \label{eq:M_AB_FFT}
\end{align}
where we have defined $\tilde \vT_\alpha = \vF ~ \vT_\alpha$ and 
$\tilde \vC_{R~I}^{-1} = \vF^{-t} ~ \vC_{R~I}^{-1} ~ \vF^{-1}$.
Note that while $\tilde \vC_{R~I}^{-1}$ is a 2D (inverse) Fourier transform, 
$\tilde \vT_\alpha$ is a 1D Fourier transform, with each column of the matrix
containing the 1D Fourier transform of each eigenvector.

\begin{figure}[H]
 \begin{center}
  \includegraphics[scale=0.5]{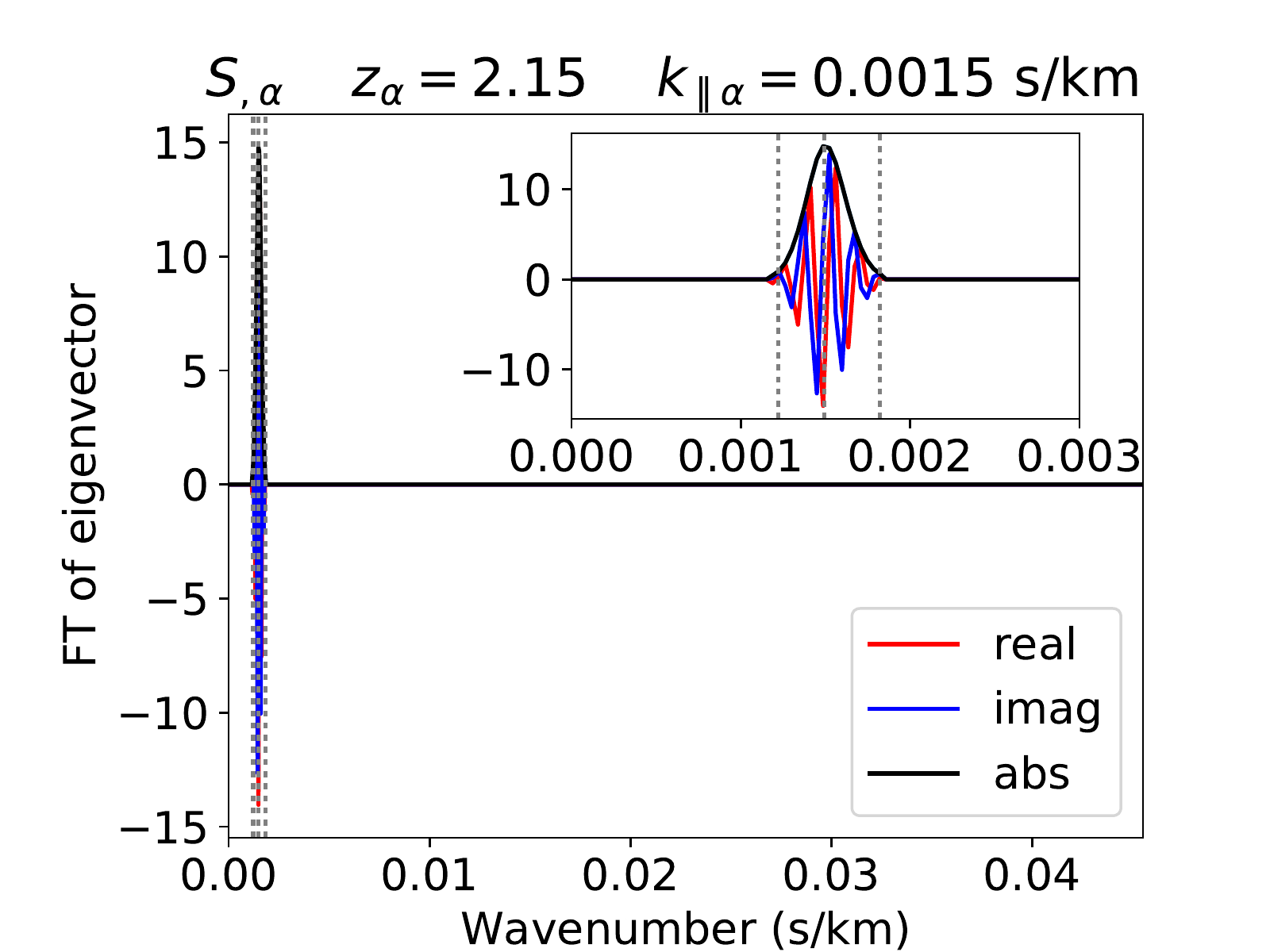}
  \includegraphics[scale=0.5]{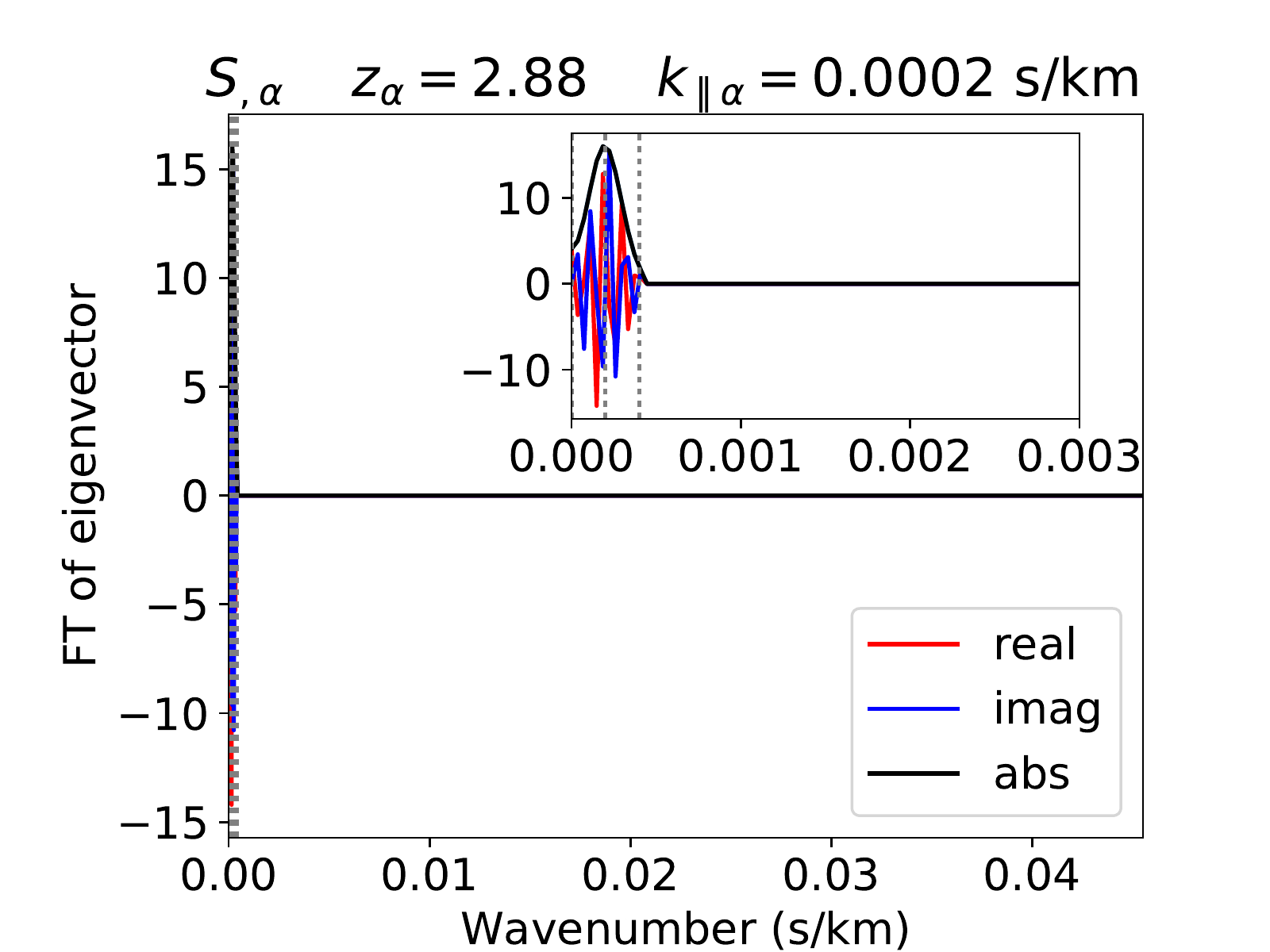}
 \end{center}
 \caption{Fourier transform of the eigenvectors shown in Figure \ref{fig:PLA_T},
  from the decomposition of the $\tilde \vS_{,\alpha}$ matrices shown in 
  Figure \ref{fig:dSdp}.
  The vertical dotted lines show the center and the edges of the wavenumber bin
  $k_{\parallel\alpha}$ for the parameter.
  The inner sub-plot zooms in to a shorter wavenumber range to better show 
  the very compact eigenvector in Fourier space.
 }
 \label{fig:PLA_FT}
\end{figure}

In Figure \ref{fig:PLA_FT} we show the Fourier transform of the eigenvectors
shown in figure \ref{fig:PLA_T}. 
It is the sparsity of these modes what results in another big speed up in the 
computation of the different $\vM_{I\alpha\beta}$ objects. 

The modes for some eigenvectors with negative eigenvalues are not as 
localized as those shown in figure \ref{fig:PLA_FT}, and for this reason 
we apply a cut in the tails of the Fourier modes, and we only keep 
$f_{\rm norm}=0.999$ of its norm.

In order to speed up the computation of equation \ref{eq:M_AB_FFT}, we  
use another approximation. 
In the limit of infinitely long windows, the matrices $\tilde \vC_{R~I}^{-1}$
would be diagonal.
We checked that including off-diagonal correlations between modes separated by 
less than $\Delta k_{\rm off} = 5.57 \times 10^{-4} \ikms$ (corresponding to 
15 off-diagonal elements in our default setting) was enough to recover unbiased
results in the analysis on mock data sets presented in Section \ref{sec:P1D}. 

\pvmhid{Note that $\Delta k_{off}$ comes out basically the same as your 
$F_k$, surely not coincidentally. You could probably merge them into one 
parameter... (or maybe define one relative to the other, since you may 
want a little relative adjustment but generally they scale together) } 

\subsection{Summary of numerical optimizations}
\label{ss:summary}

To summarize, we are trying to evaluate equations \ref{eq:dL1} and 
\ref{eq:dL2}. 
The brute force evaluation would require us to deal with matrices of size 
$N_q N_p$ where $N_q$ is the number of quasars and $N_p$ is the number 
of pixels in a spectrum. 
We do the following approximations.
\begin{itemize}
\item Expand around the model with zero cross-power, in effect
  reducing the covariance matrix to $N_q$ dense matrices of size $N_p$. 
  All relevant terms can now be calculated by
  manipulating (millions of) $N_p \times N_p$ matrices.
\item Diagonalizing the $N_p\times N_p$ matrices of derivatives of
  covariance matrix with respect to clustering parameters. 
  This allows us to project relevant eigenvectors separately for each 
  spectrum, 
  speeding up consequent operations.
\item Computing the projections of the weighted data vector and more 
  importantly inverse covariance matrix ($\vy_{I\alpha}$ and 
  $\vM_{I\alpha\beta}$) by rotating to Fourier space where the signal 
  eigenvectors become very sparse, and $\tilde{\vC}_{R~I}^{-1}$ fairly 
  diagonal, then rotating back.
\end{itemize}

There are several parameters that control the accuracy of the latter two
approximations, which we summarize in Table \ref{tab:knobs}.
\begin{table}[H]
  \centering
  \begin{tabular}{ccp{13cm}}
   Parameter & Default value & Meaning \\
\hline
   $e_{\rm min}$ & 0.02 & Keep eigenmodes whose absolute eigenvalue is
                          larger that $e_{\rm min}$ of maximum eigenvalue.\\
   $f_{\rm norm}$ & 0.999 & Cut tails of eigenvector FFTs so that it contains
                          $f_{\rm norm}$ of its norm.\\
   $\Delta k_{\rm off}$ & $5.57 \times 10^{-4}$ & Ignore correlations in 
                          $\tilde \vC_{R~I}^{-1}$ between modes separated by 
                          more than $\Delta k_{\rm off}$ (in $\ikms$).\\
   $F_z$ & 0.175 & Ignore elements of Fisher matrix ($F_{\alpha\beta}$) with 
                          $ | \Delta \ln{(1+z)} | > F_z$.\\
   $F_k$ & 0.0005 & Ignore elements of Fisher matrix ($F_{\alpha\beta}$) with
                          $ | \Delta \kpar | > F_k$ (in $\ikms$). \\
  \end{tabular}
  \caption{Parameters controlling the accuracy of our approximations.  }
  \label{tab:knobs}
\end{table}

\section{Analysis of the 1D power spectrum on mock data}
\label{sec:P1D}

We will begin our numerical exploration by using mock datasets to
compare measurements of the 1D power spectrum $P_{1D}(z,\kpar)$
implemented using brute force real space matrix multiplication 
(as done in, e.g., \cite{2006ApJS..163...80M}) 
with our numerical implementation presented
in \ref{sec:numeric}. The main motivation here is to establish basic
sanity of our numerical work and to test the validity of
approximations that we describe in Section
\ref{sec:numeric}. Measurements of 1D power spectrum alone are
sufficiently fast that we can afford to switch these approximations
off and see the impact that they have.

\subsection{Mock dataset for $P_{1D}(z,\kpar)$ measurements}

We simulate data mimicking the quasar spectra from the twelfth data 
release (DR12, \cite{2015ApJS..219...12A}) of the Sloan Digital Sky Survey 
(SDSS-III, \cite{2011AJ....142...72E}), containing mostly spectra from the 
Baryon Oscillation Spectroscopic Survey (BOSS, \cite{2013AJ....145...10D}).
We generate mock spectra for 181,506 quasar spectra with $z>2.1$, using 
quasar continua matching the amplitude of each observed quasar, and adding 
Gaussian noise to the spectra using the noise variance estimated by the 
SDSS pipeline.

The mock spectra have a mean transmitted flux fraction $\bar F(z)$ that 
approximately matches the one measured in the data, and Gaussian fluctuations 
around this mean are generated using an analytical expression for 
$P_{1D}(z,\kpar)$ from \cite{2013A&A...559A..85P}, 
corrected to be flat at very low-k.
A continuum for each spectrum is generated using a mean restframe shape for 
all quasars, redshifted and normalized accordingly to roughly match the 
observed magnitude.
Finally, the spectra are convolved using the mean resolution within each 
forest, pixelized using the pixel width of the SDSS coadded spectra, 
$T = 69.03 \kms$, and Gaussian noise is added using the variance provided
by the SDSS pipeline.

\subsection{Band power measurement}

We parameterize $P_{1D}(z,\kpar)$ following equation \ref{eq:Px_param}, for the
special case of $\Dtheta=0$, and we use the same $P_{1D}^{\rm fid}(z,\kpar)$
that was used to generate the mock spectra. 
\footnote{We have tested that we recover the right power spectrum when
starting with a fiducial not far from the truth.}
We will use a grid of band power parameters with $N_z = 7$ \textit{z-bins} 
and $N_k = 21$ \textit{k-bins}, for a total of 147 parameters.
The first z-bin is centered at $z=1.936$, and the other bins are separated by $\Delta \ln(1+z) = 0.07$.
The last z-bin is centered at $z=3.468$.
The first k-bin is centered at $\kpar=0$, and the following 5 bins 
are linearly spaced with a separation of $\Delta \kpar = 0.0002 ~ \ikms$.
The last 15 k-bins increase logarithmically with $\Delta \ln{\kpar} = 0.2$, 
with the last bin centered at $\kpar = 0.0201~\ikms$. 

\begin{figure}[H]
 \begin{center}
  \includegraphics[scale=0.45]{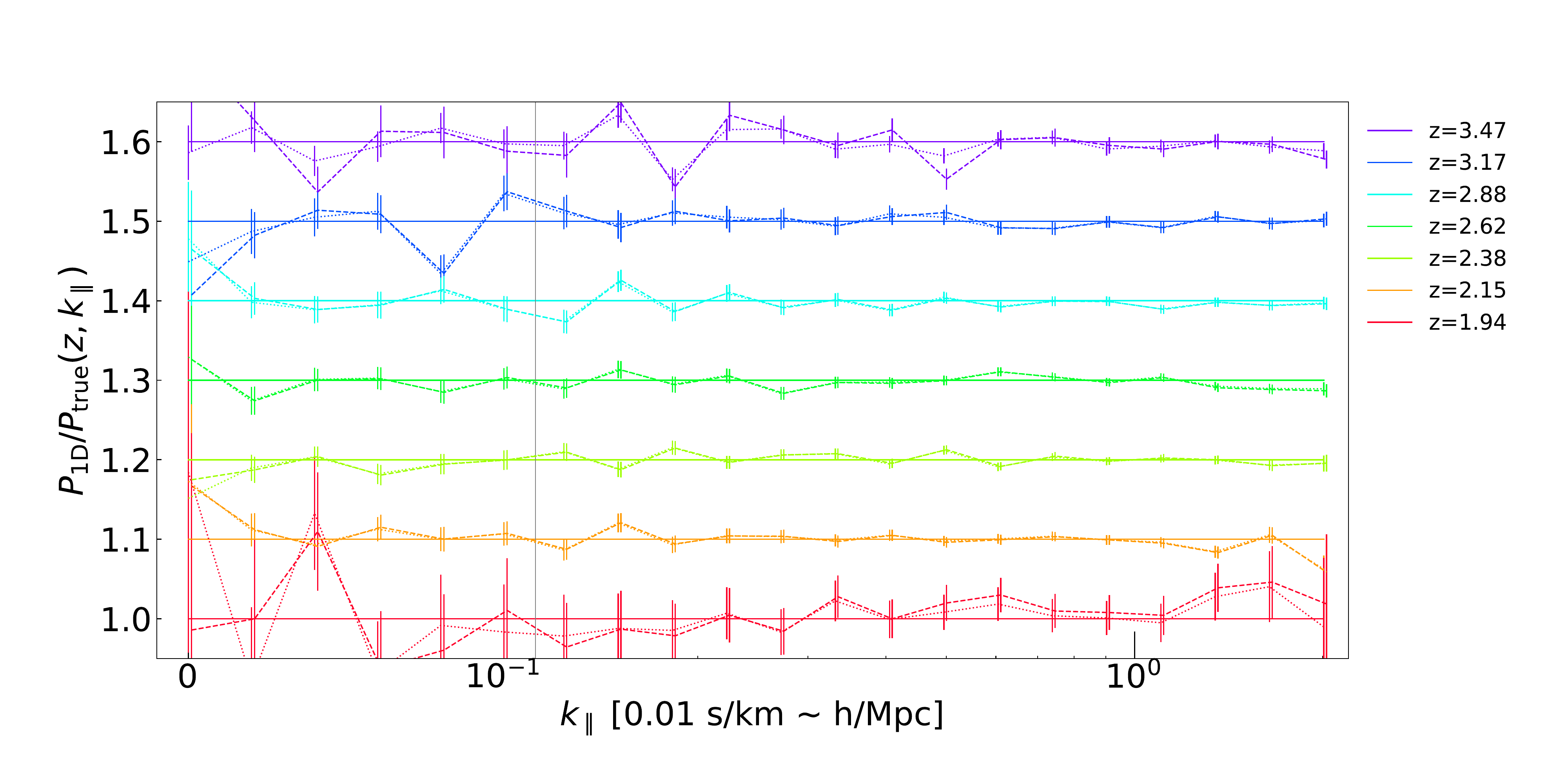}
 \end{center}
 \caption{Ratio of P1D measured over input power used to generate the mocks,
  for the different redshift bins as measured using the standard likelihood
  estimator (dotted lines) and the 
$P_\times$-oriented implementation presented in section
  \ref{sec:numeric} (dashed lines, with linear points shifted by $0.001$ and 
  logarithm points multiplied by $1.01$ for better visualization).
  The vertical line divides the linearly spaced bins (to the left) from the
  logarithmically spaced bins (to the right).
  The wavenumbers in the x-axis have been multiplied by 100 so they can be 
  approximately compared to wavenumbers in units of $\ihMpc$.
}
 \label{fig:P1D}
\end{figure}

We present the measured power in figure \ref{fig:P1D}, where we plot the 
measured power divided by the input power used to generate the mock spectra.
The definition of ``measured power'' in all of our plots is the maximum 
likelihood 
implied by our Taylor expansion of the likelihood, equivalent to the result  
of a single NR step given by Equation
\ref{eq:NR}. The errorbars are the square root of the diagonal elements
of the inverse of the Fisher matrix. 
The dotted lines show the measurement using brute force matrix 
multiplication, while the dashed lines shows the measurement
when using the $P_\times$-oriented 
implementation described in section \ref{sec:numeric}.
We see that both methods agree well with each other and the expectation for 
these mocks, with the biggest deviations maybe not surprisingly appearing at
the extremes of $k$ and $z$. 

\subsection{Goodness of fit - $\chi^2$}

In the $P_{1D}(z,\kpar)$ analysis above we computed the Taylor expansion of 
the log-likelihood around the true power spectrum $P_{1D}^{\rm fid}(z,\kpar)$.
In this case, the expected value of the first derivatives is zero, and their
covariance equals the Fisher matrix, 
$\left< \sL_{,\alpha} \sL_{,\beta} \right> = F_{\alpha \beta}$.
The standard $\chi^2$ can then be computed directly from the derivatives using
$\chi^2 = \sL_{,\alpha} F^{-1}_{\alpha \beta}  \sL_{,\beta}$
and its expected value is equal to the number of band power parameters.

In table \ref{tab:P1D_chi2} we show the performance of the different 
algorithms, and the sensitivity to the different approximations discussed 
in Section \ref{sec:numeric}.
When using all the (147) bins in the measurement, we measure a value of 
$\chi^2=155.1$ for the exact matrix multiplication analysis 
(probability of 30.7\%), 
and a value of $\chi^2=176.9$ for our new default implementation 
(probability of 4.7\%). 
$\kpar=0.02 \ikms$ is the limit typically used for data of this resolution, 
since on large scales the suppression of power caused by the spectral 
resolution is very large and difficult to correct for accurately.
If we marginalize over the last $\kpar$ bin at $k=0.0201 \ikms$, the agreement 
with the new implementation is considerably improved to $\chi^2=156.0$ for 
140 degrees of freedom (d.o.f), and the probability is of 16.9\%. 
In Appendix \ref{sec:margin} we describe how we marginalize the 
likelihood over unwanted parameters.

In table \ref{tab:P1D_chi2} we also show the effect of the different numerical 
settings on the values of $\chi^2$ as a function of $k_{\parallel ~ \rm max}$. 
\begin{table}[H]
  \centering
  \begin{tabular}{|c|c c|c c|c c|}
   \hline
    Analysis & \multicolumn{6}{c|}{$\chi^2$, prob} \\
   \hline
    $k_{\parallel ~ \rm max} [\ikms]$ (d.o.f.) 
          & \multicolumn{2}{c|}{$0.0201 ~ (147)$} 
          & \multicolumn{2}{c|}{$0.0164 ~ (140)$} 
          & \multicolumn{2}{c|}{$0.0135 ~ (133)$} \\
   \hline
    Default settings & 176.9 & 0.047 & 156.0 & 0.169 & 139.9 & 0.323 \\
    $F_z=0.1$ & 176.9 & 0.047 & 156.0 & 0.169 & 139.9 & 0.323 \\
    %$F_z=0.05$ & 177.0 & 0.046 & 156.1 & 0.167 & 140.1 & 0.321 \\
    $F_k=0.00045$ & 176.9 & 0.047 & 156.0 & 0.168 & 140.0 & 0.322 \\
    %$F_k=0.00041$ & 177.2 & 0.045 & 156.2 & 0.165 & 140.2 & 0.317 \\
    %$F_k=0.00040$ & 234.6 & 5.8e-6 & 213.7 & 6.1e-5 & 197.6 & 2.4e-4 \\
    $e_{\rm min}=0.03$ & 177.4 & 0.044 & 156.4 & 0.162 & 140.3 & 0.316 \\
    $f_{\rm norm}=0.995$ & 184.4 & 0.020 & 163.5 & 0.085 & 144.2 & 0.239 \\
    $\Delta k_{\rm off}=7.43 \times 10^{-4} \ikms$ & 171.5 & 0.081 & 152.0 & 0.230 & 136.9 & 0.390 \\
   \hline
    exact matrix ops & 155.1 & 0.307 & 139.0 & 0.508 & 125.3 & 0.670 \\
   \hline
  \end{tabular}
 \caption{For a single mock BOSS data set, 
$\chi^2$ and probability for the different $P_{1D}$ analyses, 
  when using different maximum values for $\kpar$.
  \pvmhid{Note that the dependence may be sharp because of $\Delta k_{off}$,
     which cuts mode correlations out of the calculation anyway. 
     This is why it would be good 
     to have a baseline parameter that scales both of them together (fine for
     $F_k$ here to play that role), then another parameter that allow you to
     explore varying the $\vC$ cut relative to that.}
  The default analysis uses $F_z=0.175$, $F_k=0.0005$, $e_{\rm min}=0.02$, 
  $f_{\rm norm}=0.999$ and $\Delta k_{\rm off}=5.57 \times 10^{-4} \ikms$. 
  Note that the only purpose of this analysis is to help understand the 3D 
  algorithm -- 
  we would not use this algorithm if we were only interested in 1D power. 
 }
 \label{tab:P1D_chi2}
\end{table}
One can see that we are not very sensitive to the details of the numerical
implementation, although $\Delta k_{\rm off}$ does seem to make a 
non-negligible difference. 
Our algorithm is not optimized for $P_{1D}$, so we take these results as good
enough and move on to our real goal of measuring 3D cross-correlations.

\section{Analysis of the cross-power on mock data}
\label{sec:PX}

After having established validity of the method on 1D power spectrum
measurement, we proceed to cross-power spectra.

\subsection{Mock dataset for $P_\times(z,\Dtheta,\kpar)$ measurements}

We will present the measurement on 40 realizations of the 
BOSS DR11 mocks, that have been widely used in past BOSS \lya\ BAO 
analyses \cite{2015A&A...574A..59D,2017A&A...603A..12B}.
These mocks were generated for the eleventh data release (DR11) of BOSS, for a 
total of 134 386 quasars.
Since BOSS did not generate mock spectra for the final data release (DR12),
and since the difference in area is roughly 10\%, we will use these mocks 
as realistic simulations of the BOSS survey.

The algorithm to generate the mock \lyaf\ skewers is described in
detail in \cite{2012JCAP...01..001F}.  In short, a correlated and
anisotropic Gaussian field is computed along all lines of sight, with
an input power spectrum such that after applying a lognormal
transformation to the Gaussian field to obtain an optical depth, the
resulting field has the desired \lyaf\ power spectrum from
\cite{2003ApJ...585...34M}. The important thing to note is that while
these mocks have been extensively used in the past BOSS paper, they
are known to contain imperfections, especially on small scales. Some
of these imperfections are related to redshift interpolation applied
when making these mocks and would have been missed for analysis that
use a single bin across the survey.

\subsection{Band power measurements}

We will use the same grid of redshift ($N_z = 7$) and line of sight 
wavenumbers ($N_k = 21$) bins that were used in measuring $P_{1D}(z,\kpar)$
in section \ref{sec:P1D}, and we will use a total of $N_\theta = 27$ 
angular separation bins: the first bin will be at zero separation (containing
1D information), followed by five bins linearly spaced by an
angular separation of $0.0002$ radians, or roughly $1.2$ transverse $\Mpc$; 
the last 21 bins are distributed logarithmically with a fractional increase 
of $0.2$, with a last bin centered at $0.0606$ radians, or roughly $360 \Mpc$.

\begin{figure}[H]
 \begin{center}
  \includegraphics[scale=0.35]{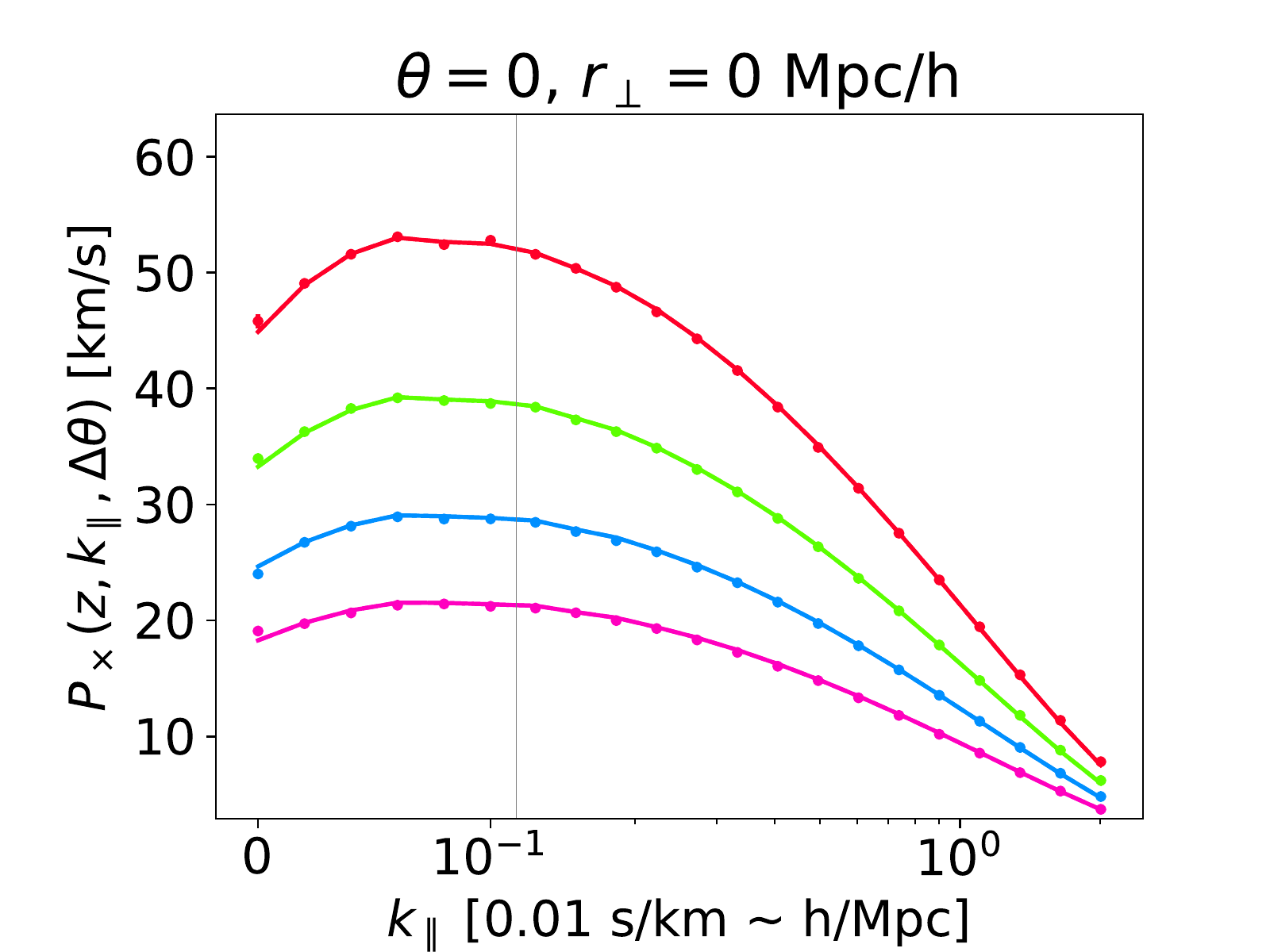}
  \includegraphics[scale=0.35]{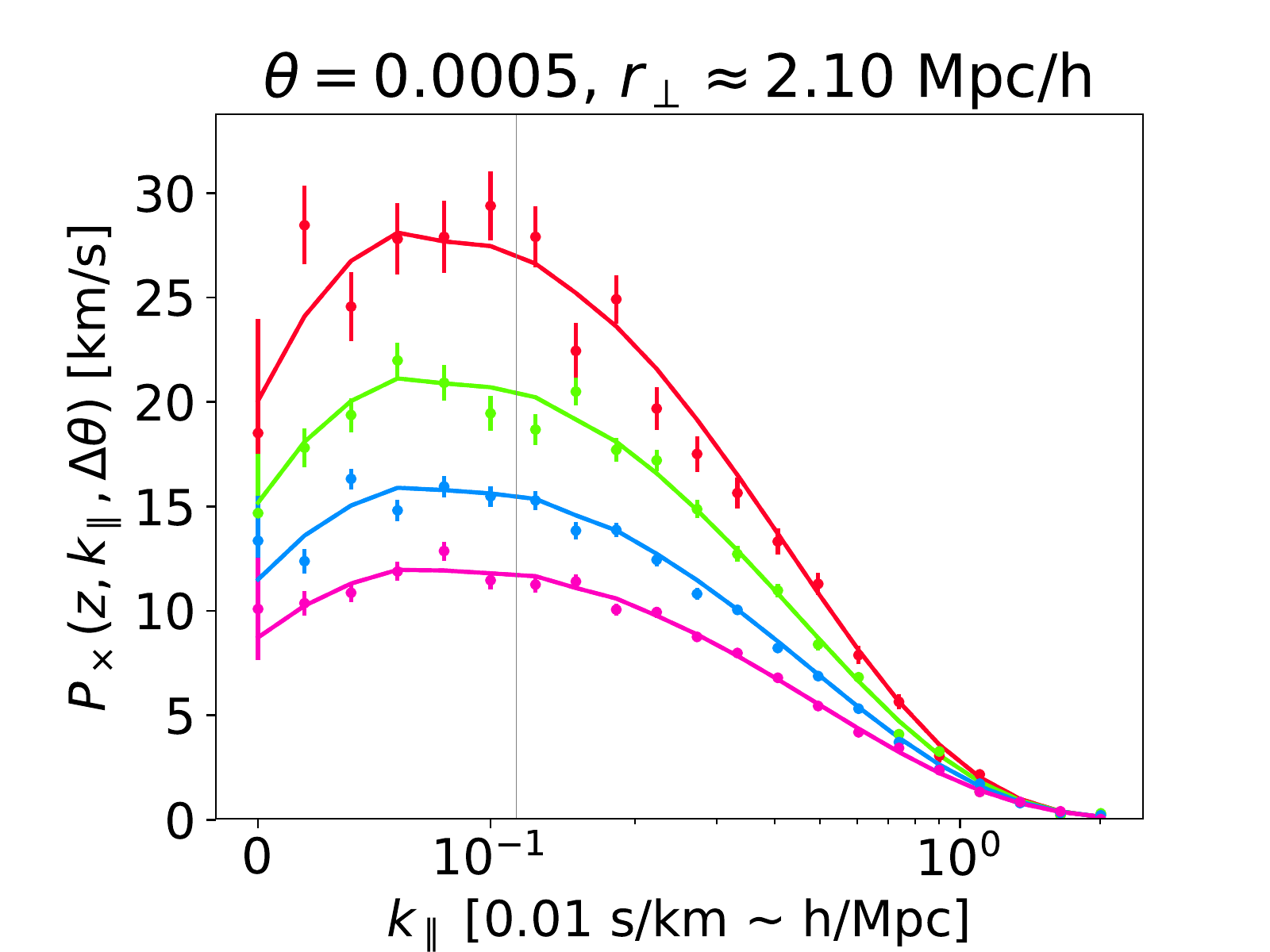}
  \includegraphics[scale=0.35]{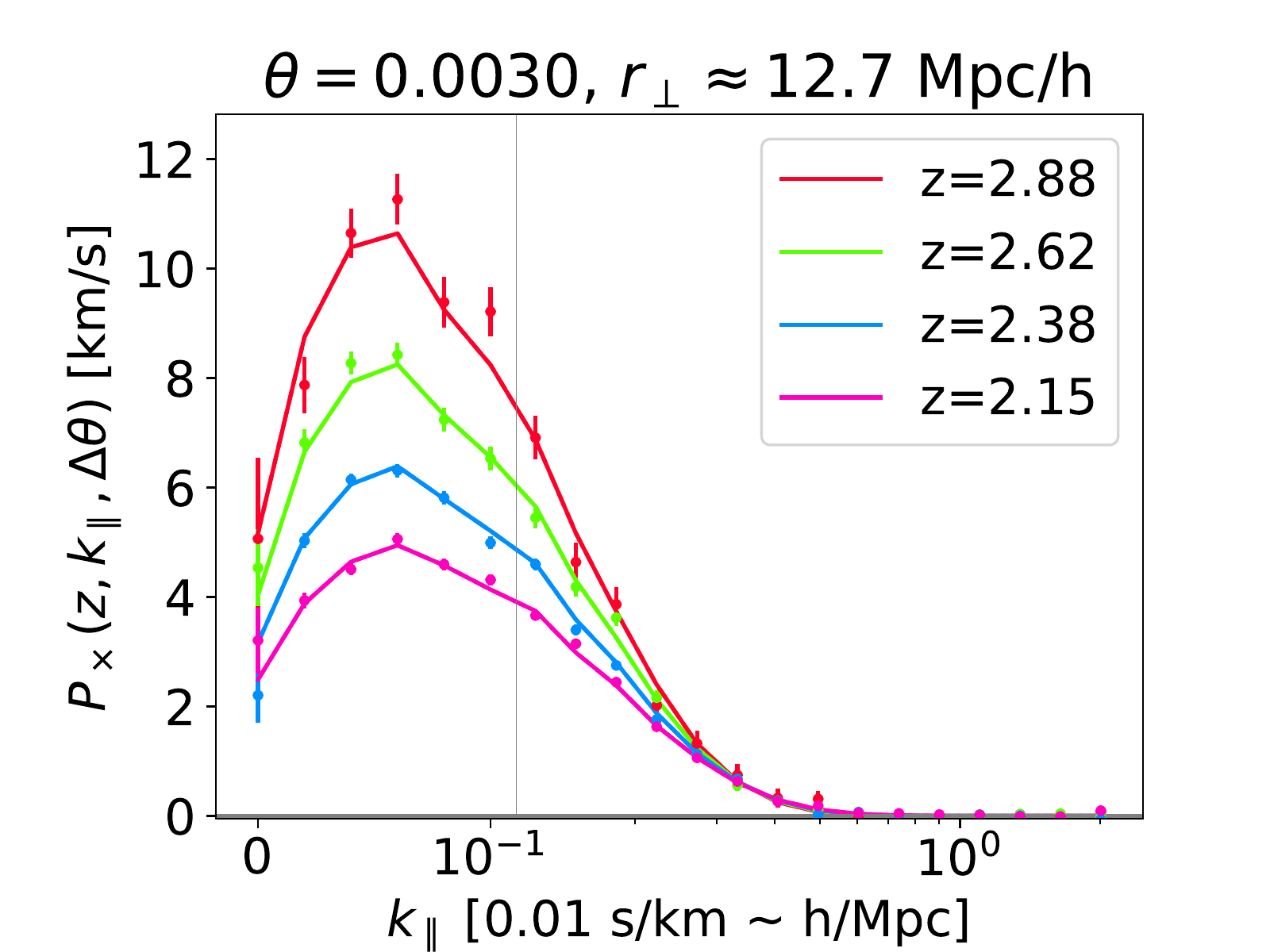}
 \end{center}
 \caption{$P_\times(z,\Dtheta,\kpar)$ measured from 40 mock realizations 
  of the BOSS survey, as a function of wavenumber $k_\parallel$, with different
  colors identifying different redshift bins, and each panel showing a 
  different angular separation bin.
  We only show the four redshift bins that are better measured, and a 
  representative sample of angular separations.
  The wavenumbers in the x-axis have been multiplied by 100 so they can be
  approximately compared to wavenumbers in units of $\ihMpc$.
  In the title of each plot we have approximated the comoving transverse 
  separation as $r_\perp \approx 4200~\theta \hMpc$.
  The vertical line divides the linearly spaced bins (to the left) from the
  logarithmically spaced bins (to the right).
 }
 \label{fig:PX}
\end{figure}

In Figure \ref{fig:PX} we show the cross-power spectrum 
$P_\times(z,\Dtheta,\kpar)$ measured from 40 mock realizations of the BOSS
survey.
The measured power is as usual plotted at the maximum likelihood point 
implied by our quadratic expansion of $\ln L$, i.e., the result of a single
NR step away from $\vp_0$ (Equation
\ref{eq:NR}). The errorbars are the square root of the diagonal elements
of the inverse of the Fisher matrix.
We show a representative sample of angular separations, and the four redshift
bins that are better measured in the analysis.
The solid lines show the expected measurement given the input theory that went
into generating the mock data, described in more detail in section
\ref{ss:theory}.
While we plot results for diagnostic purposes, we do not intend $P_\times$ to 
be an end-stage result of our algorithm -- it is an intermediate data 
compression step in the process of producing $P_{\rm 3D}$.

\section{From $P_\times(z,\Dtheta,\kpar)$ to $P_{3D}(z,k_\perp,\kpar)$}
\label{sec:P3D}

In the previous sections we have presented an algorithm to obtain a second order
Taylor expansion of the log-likelihood function around a fiducial
model spectrum, describing the cross-spectrum of the 
\lyaf\ $P_\times(z,\Dtheta,\kpar)$.

For a sufficiently fine binning, this presents a complete summary
statistics describing 2-pt correlations, and we could decide to stop here. 
We could compute theoretical predictions for the cross-spectrum, and do 
parameter estimation (BAO scale, linear power...) directly in this space. 
However, since the models are usually defined in Fourier space, it is useful 
to translate the measurement into a truly three-dimensional power spectrum
$P_{3D}(z,k_\perp,\kpar)$. 
In principle we could convert the $P_\times$ measurement into any other 
parameterization, including, for example, the correlation function.

\subsection{Determining $P_{3D}(z,k_\perp,\kpar)$}

We would like to compute the Taylor expansion of the (log-) likelihood 
with respect to a new set of parameters $\ptD_\gamma$.
Using the derivative chain rule we can compute the new derivatives 
($\sL^{(3D)}_{,\gamma}$,$\sL^{(3D)}_{,\gamma \delta}$) as a function of the 
old ones ($\sL^{(\times)}_{,\alpha}$,$\sL^{(\times)}_{,\alpha \beta}$):
\begin{equation}
 \sL^{(3D)}_{,\gamma} = \sL^{(\times)}_{,\alpha} 
    ~ \frac{\partial p^\times_\alpha}{\partial p^{3D}_\gamma}
  + \sL^{(\times)}_{,\alpha\beta} 
    ~ \frac{\partial p^\times_\alpha}{\partial p^{3D}_\gamma}
		~ \Delta P_\times(z_\beta,\Dtheta_\beta,k_{\parallel_\beta})
\end{equation}
\begin{equation}
 \sL^{(3D)}_{,\gamma\delta} = \sL^{(\times)}_{,\alpha\beta} 
    ~ \frac{\partial p^\times_\alpha}{\partial p^{3D}_\gamma}
    ~ \frac{\partial p^\times_\beta}{\partial p^{3D}_\delta} ~,
\end{equation}
where 
\begin{equation}
 \Delta P_\times(z,\Dtheta,\kpar) 
  = \int \frac{d\vk_\perp}{(2\pi)^2} ~ e^{i \vDtheta \vk_\perp} 
    ~ P_{3D}^{\rm fid}(z,k_\perp,\kpar)
    - P_\times^{\rm fid}(z,\Dtheta,\kpar) ~.
\end{equation}

The second term in the first derivative takes into 
account that the Taylor expansions generally are not computed around the same 
fiducial spectrum. The reason for this complication is that for 3D
power spectrum we want to expand around the true power, while for the
cross-spectrum we are forced to set cross-power at non-zero
separations to zero for numerical expediency (as discussed in 
Section \ref{sec:numeric}).
The second derivative is not affected since its value is constant when 
assuming a Gaussian likelihood, but would have received similar
contribution had we expanded the initial likelihood to third
order. Since the two descriptions of power spectra are linearly
related, terms involving second derivatives $\partial^2 p^\times_\alpha/\partial
  p^{3D}_\gamma \partial p^{3D}_\delta$ vanish.
The definition of $\frac{\partial p^\times_\alpha}{\partial p^{3D}_\gamma}$
is clear given Eq. (\ref{eq:inverting}):
\begin{align}
 \label{eq:deriv_code} 
\frac{\partial p^\times_\alpha}{\partial p^{3D}_\gamma} &=
  \sum_\beta \left(I^{-1}\right)_{\alpha\beta} 
  \int dz~ d\vDtheta ~\frac{d\kpar}{2 \pi} ~ w_\beta(z,\Dtheta,\kpar)
  \int \frac{d\vk_\perp}{(2\pi)^2} ~ e^{i \vDtheta_\alpha \vk_\perp}
  w_\gamma(z,\vk_\perp,\kpar)   \\ \nonumber
 &\simeq 
  \int \frac{d\vk_\perp}{(2\pi)^2} ~ e^{i \vDtheta_\alpha \vk_\perp}
  w_\gamma(z_\alpha,\vk_\perp,\kpar^\alpha) 
\end{align}
where $I_{\alpha\beta}=
\int dz~ d\vDtheta ~\frac{d\kpar}{2 \pi}w_\alpha(z,\Dtheta,\kpar)
w_\beta(z,\Dtheta,\kpar)$. In Appendix \ref{sec:derivatives} we discuss in 
more detail these derivatives, and give an explicit window function linking 
3D power spectrum and 3D power spectrum parameters. 
 
We will use the same grid of ($z$, $\kpar$) parameters that was used
to measure $P_\times(z,\Dtheta,\kpar)$, i.e., a total of $N_z=7$ and 
$N_{k_\parallel}=21$ bins. 
We will define the following $k_\perp$ grid: 
the first bin will be centered at $k_\perp=0$, and it will be followed by 5 
linear spaced bins of width $\Delta k_\perp = 60$ (in units of inverse radians),
followed by a set of 15 bins logarithmically spaced 
$\Delta k_\perp / k_\perp = 0.2$, for a total of $N_{k_\perp}=21$ bins.

\begin{figure}[H]
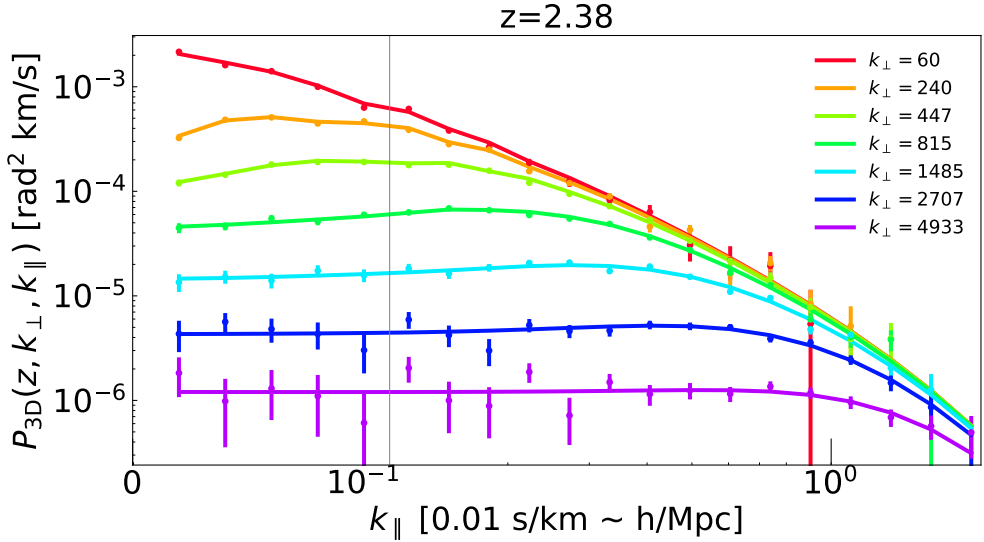

 \begin{center}
  \includegraphics[scale=0.5]{{{Pzlkp_wP1D_marg_comb039_z2.38}}}
 \end{center}
 \caption{$P_{3D}(z,k_\perp,\kpar)$ measured as a function of line of sight 
  wavenumber $\kpar$, for several different transverse wavenumbers $k_\perp$, 
  at $z=2.38$. 
  Measurement from 40 synthetic realizations of the BOSS survey. 
  The solid lines show the input theory used to generate the mocks.
  The radial wavenumbers in the x-axis have been multiplied by 100 so they 
  can be approximately compared to wavenumbers in units of $\ihMpc$.
  The vertical line divides the linearly spaced bins (to the left) from the
  logarithmically spaced bins (to the right).
 }
 \label{fig:Pzlkp}
\end{figure}

After computing the Taylor expansion of the log-likelihood with band powers
$\vp^{3D}$ we can visualize the results by computing the implied maximum of 
the resulting Gaussian 
likelihood approximation (i.e., the result of a single 
Newton-Raphson step away from $\vp_0$). 
Figure \ref{fig:Pzlkp} shows some of the estimated band power values, and the 
true power in the mocks.
Note that in figure \ref{fig:Pzlkp} we have marginalized the likelihood over 
band powers corresponding to bins with $\kpar=0$ or $k_\perp=0$, following
the recipe described in Appendix \ref{sec:margin}.
We expect these modes to be very difficult to measure in a real data analysis,
since $\kpar=0$ modes are very sensitive to distortions caused in continuum 
fitting and $k_\perp=0$ are very sensitive to systematics that vary slowly 
in the sky. 

\begin{figure}[H]
 \begin{center}
  \includegraphics[scale=0.6]{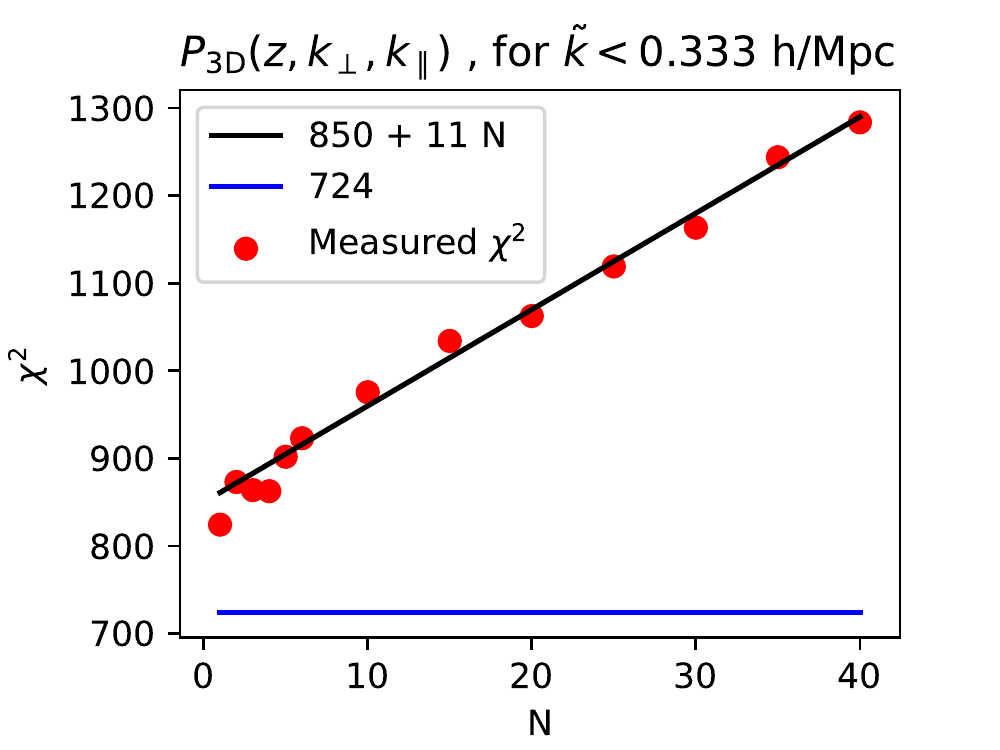}
  \includegraphics[scale=0.6]{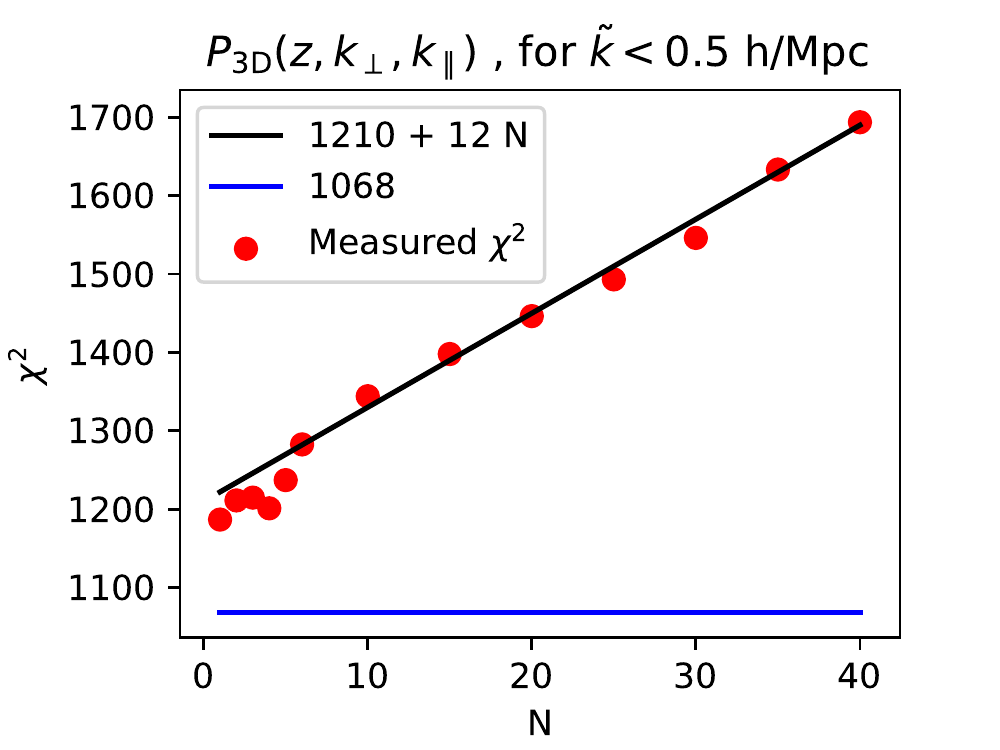}
  \includegraphics[scale=0.6]{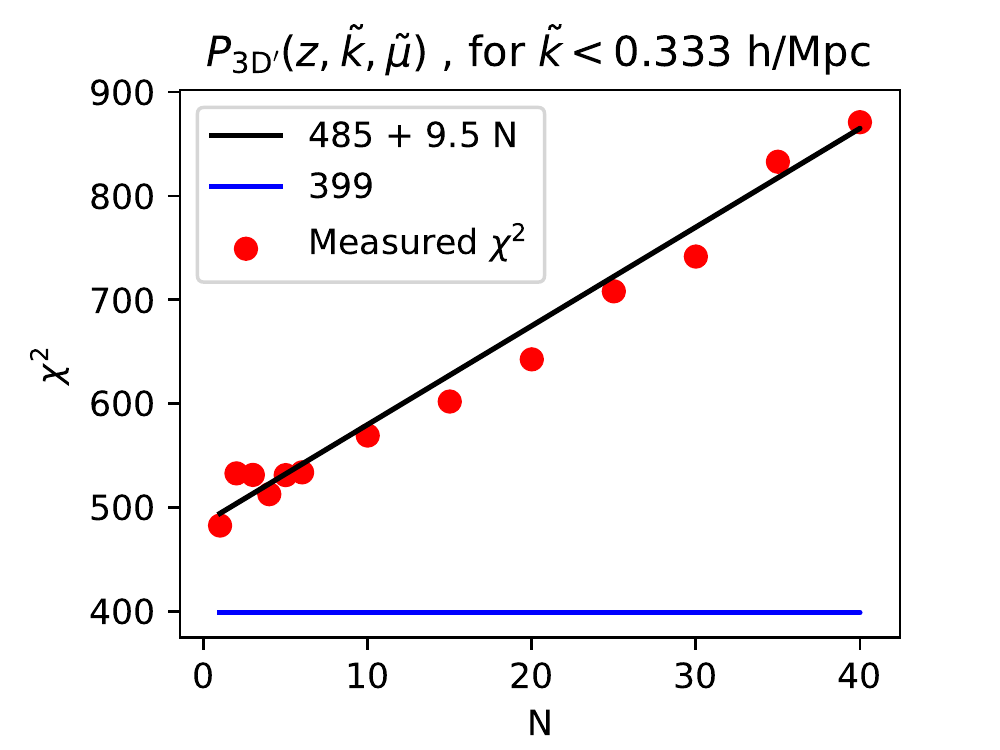}
  \includegraphics[scale=0.6]{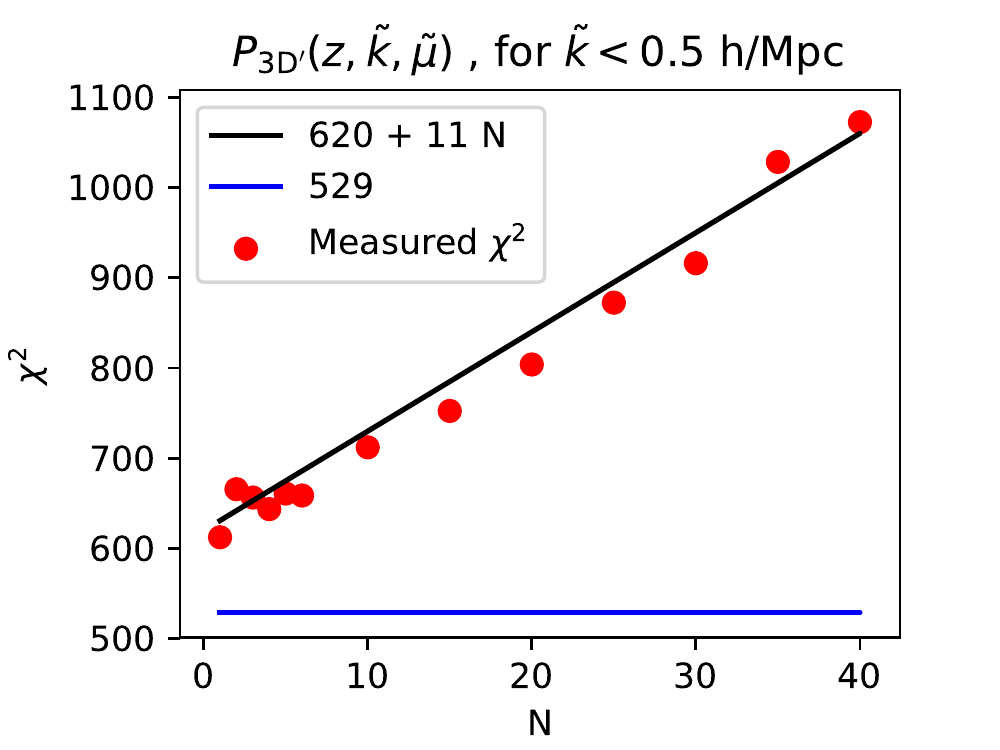}
 \end{center}
 \caption{Value of $\chi^2$ as a function of number of realizations combined,
  $N$, when comparing the measured band powers to the input model, 
  after marginalizing over bands with either $\kperp=0$ or $\kpar=0$.
  The top panels show the results for $P_{3D}(z,\kperp,\kpar)$, while the 
  bottom panels show the results for $P_{3D^\prime}(z,\tilde k,\tilde \mu)$. 
  In both cases we also marginalize over modes with 
  $\tilde k > 0.00333 \ikms \approx 0.333 \ihMpc$ (left) or
  $\tilde k > 0.005 \ikms \approx 0.5 \ihMpc$ (right). 
  The blue line shows the number of degrees of freedom, and the black line is 
  an approximate fit to the linear dependence of $\chi^2$ with the number 
  of realizations (see text).
 }
 \label{fig:Pzlkp_chi2_N}
\end{figure}

In the bottom panels of Figure \ref{fig:Pzlkp_chi2_N} we measure our success 
quantitatively by computing $\chi^2$ for the estimated power relative to 
our expectation for these mocks. 
We see that $\chi^2$ per degree of freedom is near 1 but slightly higher. 
The value increases with the number of realizations averaged together, 
suggesting some systematic errors that are becoming larger relative to the 
statistical errors. The $N=0$ intercept should give an indication of how 
much we have underestimated our statistical errors, i.e., 
$\chi^2/\nu=1.13-1.17$ indicating 6-8\% underestimation of rms errors 
(depending on the maximum $k$ used). In general the underestimation could
come from neglected non-Gaussianity in the field, expanding around zero
cross-correlation, or errors in the various other approximations in our 
algorithm. Qualitatively, these results seem pretty good, but the more 
important question is how these errors propagate into measurement of 
more relevant parameters (like the BAO scale or the amplitude of the linear 
power spectrum).

It is not always trivial to compute a value of $\chi^2$ from the Taylor 
expansion of the likelihood, since often the Fisher matrix is not positive
definite. 
For instance, this is clearly the case whenever we have more 
$\ptD_\gamma$ than $\px_\alpha$ parameters.\footnote{To be clear, a singular Fisher matrix is not a problem for our 
main science goal of propagating the information in the data into final 
cosmological parameter constraints. 
In fact, if not for computational limitations we would ideally use an
essentially infinite number of parameters, i.e., a continuous parameterization
of our function. The parameters would not be individually constrained in 
this limit, but would provide the best possible representation of the
information we have.}
In general, in order to prevent poorly measured modes to contribute to the 
value of $\chi^2$, we compute an eigenvalue decomposition of the Fisher matrix
and consider only the contribution from those eigenmodes with an eigenvalue 
that is at least $10^{-10}$ times the largest eigenvalue.\footnote{We have tested that the exact cut used does not qualitatively affect 
the main results of this study.}
This explains why the number of degrees of freedom in this section are 
usually smaller than the number of band powers measured.

In the bottom panels of Figure \ref{fig:Pzlkp_chi2_N} we show a similar 
$\chi^2$ comparison after converting the measured power into 
$(\tilde{k},\tilde{\mu})$ bands, as described in Section \ref{ss:Pzkmu} below. 
We find a similar conclusion that rms errors are probably underestimated 
by 8-10\%.
%\begin{figure}[H]
% \begin{center}
%  \includegraphics[scale=0.5]{chi2_Pzkmu_k1000}
%  \includegraphics[scale=0.5]{chi2_Pzkmu_k1500}
% \end{center}
% \caption{Same as Figure \ref{fig:Pzlkp_chi2_N}, but for 
%  $P_{3D^\prime}(z,\tilde k,\tilde \mu)$.  }
% \label{fig:Pzkmu_chi2_N}
%\end{figure}

\subsection{Fitting bias parameters on measured $P_{3D}(z,k_\perp,\kpar)$}
\label{ss:theory}

The model describing the \lyaf\ correlations in the mocks is better described
as a function of comoving wavenumber $\vq$, related to our observed wavenumbers
by:
\begin{equation}
 \label{eq:comov_coord}
 q_\parallel = \frac{H(z)}{1+z} ~k_\parallel ~,     \qquad
 q_\perp = \frac{1}{d_A(z)(1+z)} ~ k_\perp ~,
\end{equation} 
where $H(z)$ is the Hubble parameter at the redshift where the power spectrum
is measured, and $d_A(z)$ is the angular diameter distance to that redshift.

Using these coordinates, the power measured in the mocks can be modeled by:
\begin{equation}
 P_{3D}(z,\vq) = b^2(z) ~ \left[1 + \beta(z) \mu_q^2 \right]^2 
        ~ P_L(z,q) ~ D_{NL}(q,\mu_q) ~,
 \label{eq:P3D_DR11_model}
\end{equation}
where $b(z)$ is the linear density bias, $\beta$ is the linear redshift space
distortion parameter, $q$ is the modulus of the comoving wavenumber, 
$\mu_q$ is the cosine of its angle with respect to the line of sight, 
$P_L(z,q)$ is the linear power spectrum and $D_{NL}(q,\mu_q)$ is an analytical 
expression to account for non-linearities in the flux power spectrum derived in 
\cite{2003ApJ...585...34M}.
The mocks where generated with a constant value of $\beta=1.4$, and with an
evolving bias of $b = b_0 \left[(1+z)/(1+z_0)\right]^{2.9}$, with 
$z_0=2.25$ and $b_0=-0.14$. 

If we were to fit these values from a measurement on real data (where the true
underlying model is unknown) we would need to fit at the same time the relation
between observed and comoving coordinates, i.e., $H(z)$ and $d_A(z)$. 
However, in this section we will fit $b_0$ and $\beta$ on the measured 
likelihood expansion while keeping fixed the rest of the model, and using 
the conversion factors from the cosmological model used in generating the mocks.

For each pair of values for ($b_0$,$\beta$) we compute the prediction for
all band power parameters $p_{3D}$ by simply evaluating the model
power at the central $\vk$ and $z$ for the parameter (if we wanted to be
more careful we could use the functional inversion formula, 
Eq. \ref{eq:inverting}, again to define the conversion). 
We find the best fit parameters that maximize the likelihood (the standard
procedure for this final small number of non-linear parameters, rather than
continuing to use the chain rule which assumes the resulting likelihood is
sufficiently Gaussian). 
We get uncertainties in these parameters by looking at the second derivative
of the log-likelihood evaluated at the best fit parameters. 
In table \ref{tab:fitPzlkp} we show the best fit values of $b_0$ and $\beta$
for two different analyses: including $P_{1D}(z,\kpar)$ information, and
marginalizing over it.

\begin{table}[H]
 \center
 \begin{tabular}{c | c | c c | c}
  Include $P_{1D}$ & $100~\tilde k_{\rm max} [\ikms] $ & $b_0 ~ (=-0.14)$ & 
                                $\beta ~ (=1.4)$ & $\chi^2$ (d.o.f.) \\
 \hline
  Y & 2.00 & $-0.1392 \pm 0.00004$ & $1.412 \pm 0.0011$ & 6361.3 (2473) \\
  Y & 1.00 & $-0.1398 \pm 0.00018$ & $1.404 \pm 0.0039$ & 2497.0 (1800) \\
  Y & 0.50 & $-0.1398 \pm 0.00020$ & $1.406 \pm 0.0043$ & 1693.8 (1068) \\
 \hline
  N & 2.00 & $-0.1398 \pm 0.00019$ & $1.403 \pm 0.0040$ & 3604.8 (2631) \\
  N & 1.00 & $-0.1399 \pm 0.00019$ & $1.402 \pm 0.0041$ & 2670.4 (1943) \\
  N & 0.50 & $-0.1398 \pm 0.00020$ & $1.406 \pm 0.0044$ & 1820.0 (1203) \\
 \hline
 \hline
  Y & 2.00 & $-0.1392 \pm 0.00004$ & $1.411 \pm 0.0011$ & 4817.3 (959) \\
  Y & 1.00 & $-0.1398 \pm 0.00018$ & $1.404 \pm 0.0039$ & 1416.1 (761) \\
  Y & 0.50 & $-0.1397 \pm 0.00021$ & $1.408 \pm 0.0046$ & 1072.5 (529) \\
 \hline
  N & 2.00 & $-0.1397 \pm 0.00020$ & $1.405 \pm 0.0043$ & 1994.5 (1048) \\
  N & 1.00 & $-0.1398 \pm 0.00020$ & $1.404 \pm 0.0044$ & 1538.3 (846) \\
  N & 0.50 & $-0.1397 \pm 0.00022$ & $1.408 \pm 0.0048$ & 1177.4 (603) \\
 \end{tabular}
 \caption{Best fit values of bias ($b_0$) and redshift space distortion 
  parameter ($\beta$) from the measured $P_{3D}(z,k_\perp,\kpar)$ 
  (top six rows) and $P_{3D^\prime}(z,\tilde k, \tilde \mu)$ (bottom six rows),
  combining 40 realizations of the BOSS survey. 
  Some rows included $P_{1D}(z,\kpar)$ information, while other do not. 
  The second column specifies the maximum wavenumber included in the fits, in 
  units of $\ikms$ multiplied by 100 so that the numbers can be approximately
  compared to $\ihMpc$.
  The last column shows the value $\chi^2$ with respect to the
  expected theory in the mocks (not the fit).
  Note that the value of $\chi^2$ in the third (ninth) row correspond 
  to the last point in the upper (lower) right panel of Figure \ref{fig:Pzlkp_chi2_N}. 
 }
 \label{tab:fitPzlkp}
\end{table}

In all cases, we marginalize over $\kpar=0$ and $k_\perp=0$.
For each type of analysis we show the results as a function of maximum 
wavenumber used in the fits $\tk_{\rm max}$, where $\tk$ is defined 
by Eq. \ref{eq:tildek}, 
with $f(z)=H(z)d_A(z)$ for the fiducial model, i.e., even though
our bands are labeled by $(\kpar,~k_\perp)$ here, we make the cutoff uniform
in magnitude of $\tk$. The bottom line is that, even for statistical errors
corresponding to $\sim 40$ times BOSS, there is no detectable bias in our 
measurement of the bias parameters (except in the unrealistic case where we
fit to high $k$ using 1D power). Apparently the systematic errors that produce
poor $\chi^2$ in this limit have a peculiar form that does not 
project onto these parameters (it is hard to identify these errors by eye 
given so many precisely measured points). For this large number of mocks, the
$\sim 8$\% underestimate of the statistical errors makes a sub-dominant 
contribution to the excess $\chi^2$.  

We can see in table \ref{tab:fitPzlkp} that when the $P_{1D}(z,\kpar)$ 
information is not included, the uncertainties in the bias parameters are 
fairly independent of our choice of $\tilde k_{\rm max}$. 
This means that in a BOSS-like survey, the cross-correlation between different
lines of sight mostly constrain large scales modes with 
$\tk \lesssim 0.005 \ikms$ ($q \lesssim 0.5 \ihMpc$), although radial modes 
should be well-measured to higher $k$ than transverse modes. 
This is analogous to the way that point object clustering measurements 
become shot-noise limited at small scales. 
On the other hand, when we include $P_{1D}(z,\kpar)$ the uncertainties are 
driven by $\tilde k_{\rm max}$, since we are able to measure correlations 
down to very small scales.
Similarly, by comparing the results using $\tk_{\rm max} = 0.005 \ikms$
we can see that the contribution of $P_{1D}(z,\kpar)$ is quite small, the
measurement is dominated by cross-correlations from different lines of sight.

Note that when analyzing real data we would not be able to fit the biasing
model down to these very small scales, since deviations from linear theory
are already important at $q \approx 0.2 \ihMpc$ or so (see Figure 2 of 
\cite{2015JCAP...12..017A}). 
However, the mocks we are using are well described by the model presented 
in equation \ref{eq:P3D_DR11_model} down to a fairly small scales, as shown 
by correlation function analyses by the BOSS collaboration
(\cite{2015JCAP...05..060B,2015A&A...574A..59D,2017A&A...603A..12B}).

\subsection{$\PtDp(z,\tk,\tmu)$}
\label{ss:Pzkmu}

Another example of likelihood conversion is to change to a set of 
parameters labeled by 
transformed coordinates
$\tk$ and $\tmu$ as defined in Equations \ref{eq:tildek} and
\ref{eq:tildemu}. 
Again, we use the chain rule to convert between different
representations of the power spectrum.
It is important to understand that this is really just a
different interpolation scheme to describe the same measured power spectrum
that is fundamentally a function of $(\kpar,\kperp)$. E.g., every band in 
$\PtDp(z,\tk,\tmu)$ can still be labeled by well-defined
$(\kpar,\kperp)$, it is just that we assume interpolation is linear in
$\tk$ and $\tmu$ instead of $\kpar$ and $\kperp$.
With $f(z)\equiv H(z) d_A(z)$ for a fiducial 
cosmology, $\tmu$ is approximately a proper physical angle, and $\tk$ is 
approximately the magnitude of a proper physical 
wavevector.\footnote{The measured power is still 
technically defined by the Fourier 
transform of the correlation function in observable coordinates, as in 
Eq. \ref{eq:P3Dfromxi}, so simply multiplying $\tk$ by $H(z)$ will not 
give exactly the same result one would have obtained by doing the transforms 
in comoving coordinates from the start, but this distinction is small.}
Note that in general we could go straight from $P^\times$ to 
$\PtDp(z,\tk,\tmu)$, but going through 
$\PtD(z,\kpar,\kperp)$ allows things like marginalization over $\kperp\sim 0$
which we expect may be corrupted by systematic errors. 
We compute $\PtD(z,\kpar,\kperp)$ parameters given $\PtDp(z,\tk,\tmu)$ by
simply evaluating at the central values of $(\kpar,\kperp)$, although we could
use another copy of the functional inversion Equation, (\ref{eq:inverting}),
if desired. 

\begin{figure}[H]
 \begin{center}
 \includegraphics[scale=0.5]{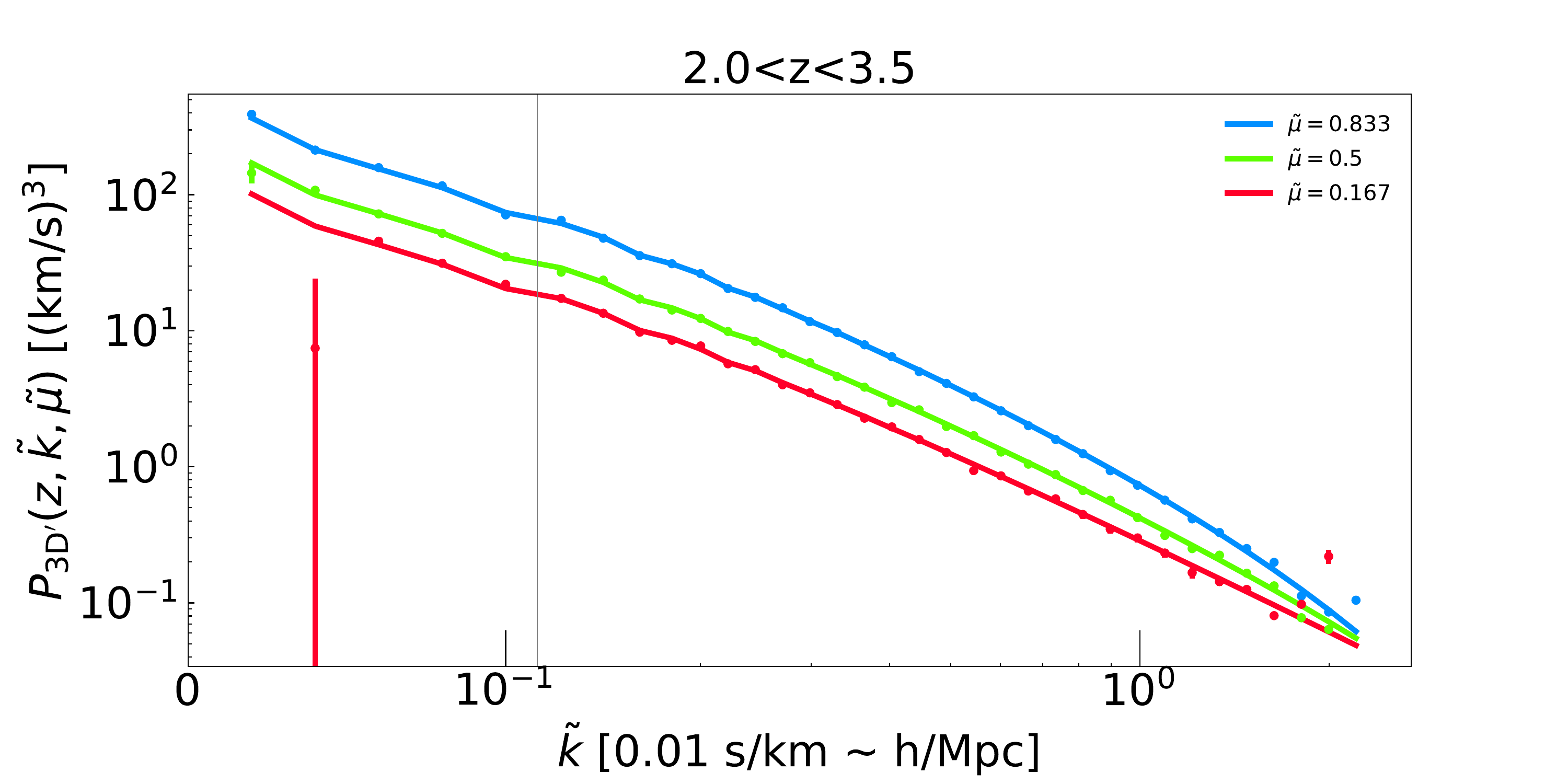}
 \end{center}
 \caption{Measured $P_{3D^\prime}(z,\tilde k,\tilde \mu)$ as a function of 
  wavenumber $\tilde k$, for different angular directions $\tilde \mu$.
  Measurement from 40 synthetic realizations of the BOSS survey.
  The solid lines show the input theory used to generate the mocks.
  The wavenumbers in the x-axis have been multiplied by 100 so they
  can be approximately compared to wavenumbers in units of $\ihMpc$.
  The vertical line divides the linearly spaced bins (to the left) from the
  logarithmically spaced bins (to the right).
  All the information has been compressed into a single redshift bin, and we 
  have used the model described in equation \ref{eq:P3D_DR11_model} as our 
  fiducial power. 
  These mocks were designed to have the correct large scale clustering for 
  BAO analysis of the \lyaf, and it is not surprising that there are 
  disagreements on the smallest scales where they had not been tested before. 
 }
 \label{fig:HighkPzkmu}
\end{figure}

In Figure \ref{fig:HighkPzkmu} we show the measured 
$P_{3D^\prime}(z,\tilde k,\tilde \mu)$, as a function of angular direction 
$\tilde \mu$. 
The plot shows the measurement from the combination of 40 realizations of the 
BOSS survey.
Whenever we change the set of parameters we can decide to use a different 
set of redshift bins, and in order to compress all the information in a single
plot in this section we plot $P_{3D^\prime}(z,\tilde k,\tilde \mu)$ measured
at a single redshift bin. 

In section \ref{ss:theory} we estimated the best fit values of the bias
parameters ($b_0$,$\beta$) from measurements of $P_{3D}(z,k_\perp,\kpar)$.
We can similarly fit for the bias parameters from measurements of 
$P_{3D^\prime}(z,\tilde k,\tilde \mu)$, as shown in the second part of 
Table \ref{tab:fitPzlkp}.  
We can see that the estimates of the bias parameters, and their uncertainties, 
are very similar to those obtained directly from $P_{3D}(z,k_\perp,\kpar)$.

\begin{figure}[H]
 \begin{center}
 \includegraphics[scale=0.45]{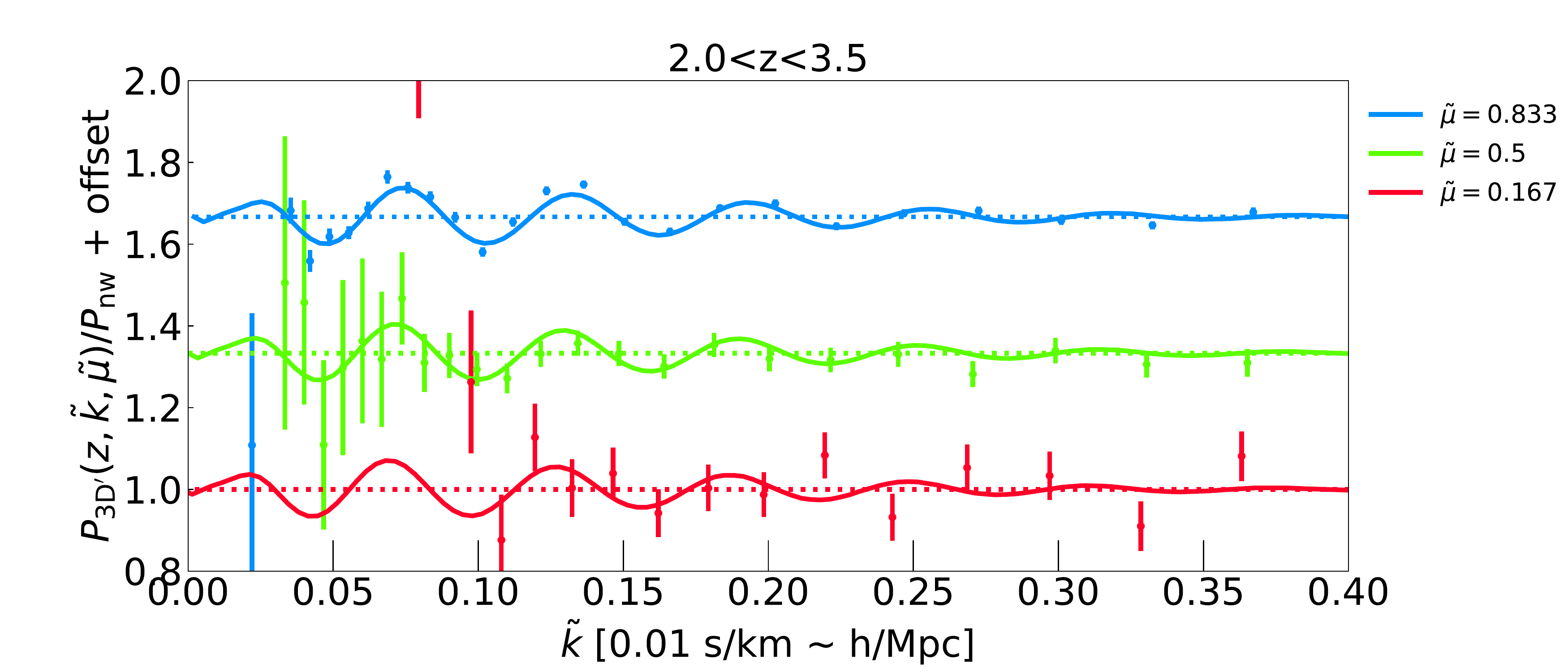}
 \end{center}
 \caption{Same $P_{3D^\prime}(z,\tilde k,\tilde \mu)$ as figure 
  \ref{fig:HighkPzkmu}, 
  but divided by a smoothed version of the power without the oscillations,  
  and zooming on larger scales relevant for BAO.
  Note that an arbitrary offset has been added for better visualization.
  Measurement on 40 mock realizations of the BOSS survey.
  In order to resolve the oscillations this analysis has used a finer binning
  in $\kpar$ and $k_\perp$ and it has ignored some of the high-k range 
  covered in figure \ref{fig:HighkPzkmu}.
 }
 \label{fig:LowkPzkmu}
\end{figure}

In order to highlight the baryonic features on large scales, in 
Figure \ref{fig:LowkPzkmu} we plot both the measured and the expected 
power spectra divided by a smooth version of the power without the 
oscillations.
In a BAO analysis it is important to resolve the oscillatory features, 
and for this reason we have used a finer binning in $\kpar$ and $k_\perp$ 
in the analysis presented in Figure \ref{fig:LowkPzkmu}.
We see, as expected, that the BAO oscillations are well-measured in the radial
direction, and much less well measured in the transverse direction. 
At this high 
precision we do see what looks like some small systematic deviations from the
curve. Presumably by the time this level of deviation matters we can tighten
up our numerics enough to eliminate it.

\section{Discussion}
\label{sec:discussion}

In this paper we have presented a new method to estimate the 3D
power spectrum of the \lyaf\ power spectrum. The central tenet
of our method is Taylor series expansion of the likelihood around a
fiducial model. We use cross-spectrum as a natural intermediate
product given the \lyaf\ data window function. The likelihood
for the cross-spectrum can be converted into likelihood for any
statistics of interest by applying chain rule to the derivatives. We have
demonstrated the applicability of this method on an idealized set of
mock data of realistic size and have justified the use of several
optimization tricks.

Finally, we discuss a few qualitative issues. 

\subsection{Coordinate system}
\label{ss:coordinates}

We have defined our power spectrum by Eq. (\ref{eq:P3Dfromxi}) to be the 
Fourier transform of the 
correlation function in observable coordinates, i.e., angle and redshift 
($\ln\lambda$), and always label our band parameters in these units.
In contrast, analyses of redshift survey data usually
use a fiducial cosmology to translate the observed coordinates 
into approximate comoving coordinates. 
Our reason for using observable coordinates is, at the core, that it is 
necessary given the algorithm that we want to use for the massive data
compression step (pixels $\rightarrow P_\times$).
We want to make an FFT or similar transform along the individual
spectra, after which we simply cannot use transverse comoving separations at
the pixel level -- we are no longer using pixels.
In contrast, if one is primarily interested in measuring statistics by
summing over pairs of galaxies or pixels, comoving separation can be
computed pair-by-pair. 
Our reason for wanting a radial transform is the usual one: 
Fourier modes at least approximately diagonalize the problem, naively
reducing the computation time by a factor of $\sim N_{\rm pixels}$ or more
(we will discuss later why our analysis of a similar set of data does not 
in fact run faster than well-developed correlation function measurements).

The important question for us is: is there anything fatally wrong with using
these coordinates? The standard objection would be, e.g., that if you measure
a single correlation function by counting pairs in angular bins over a wide
range in redshift, a given bin in angle will correspond to changing comoving
separation so a feature like BAO will be smeared. 
As long as the fiducial model is sufficiently close to correct
you can count pairs in comoving bins over an arbitrarily wide range
in redshift without blurring any features that are fixed in comoving 
coordinates, as linear regime structures like BAO are.
Our analysis, however, does not naively smear structure in this way. It is 
built from the beginning around well-defined 
interpolation in the coordinates, including measuring the power 
centered at an arbitrarily spaced set of 
points in redshift. We do not lose information by chopping the redshift range
into chunks to be analyzed independently, and we automatically compute the 
covariance matrix between parameters at different redshifts.  
For sufficiently finely spaced points, the 
interpolation can resolve any 
structure, and we can always check that it is sufficient against mocks, and
check for convergence in the real data analysis. With the linear interpolation
we use in redshift, our error is smaller than a simple estimate of smearing 
based on, e.g., the change in angular position of a fixed comoving feature
across a bin (e.g., 5.4\% between typical band centers $z=2.15$ and $z=2.38$).  
The key thing to understand is: with well-designed interpolation between 
parameters labeled by any kind of redshift and radial and transverse 
separation coordinates, the 
potential value
of comoving coordinates really boils down to providing a maximally smooth
basis for interpolation. 

There are some other reasons we might be happy to remove the layer of 
complication that comes with working in fiducial comoving coordinates:
While the large-scale (linear regime) redshift evolution 
can be captured in comoving coordinates by the evolution of two numbers,
$b_\delta(z) \sigma_8(z)$
and $f(z) b_\eta(z) \sigma_8(z)$,\footnote{In galaxy clustering analyses one
can use $b_\eta=1$, but this is not the case in analyses of the \lyaf\ 
\cite{2000ApJ...543....1M,2003ApJ...585...34M,2012JCAP...03..004S}.} 
the redshift evolution of clustering on smaller
(non-linear) scales is more complicated, and it is generally desirable to 
use relatively finely spaced redshift bins in any case to resolve this 
evolution.
Also, using a wavelength to define a redshift to define comoving coordinates
requires that you specify a transition wavelength, e.g., \lya.  
A small but non-negligible fraction of the absorption is  caused by higher
order hydrogen lines or metal
lines (predominantly C-IV, Si-II, Si-III, Si-IV, O-VI, MgII\ldots). The 
relative position of this absorption is fixed in $\ln \lambda$, and ``comoving
positions'' based on the \lya\ wavelength are simply wrong. Generally, 
observation-related systematics live naturally in angle and wavelength space
and know nothing about comoving coordinates. 
General-use spectra
typically come with pixels evenly spaced in $\ln \lambda$ so we can FFT 
in these units without any re-pixelization. 

Note that the key place where we need observable coordinates is the
pixels$\rightarrow P_\times$ compression step. After that we could easily use
the chain rule to convert to $P_{3D}$ in fiducial comoving coordinates 
(we effectively do this with $\tmu$). The potential damage of distorting
features by insufficiently accurate interpolation 
will already be done, however, so that is no longer a motivation. 
While it might seem easier to use results labeled in fiducial
comoving units, suggesting that if we published results from data 
we might want to at least shift at the end to these units, this is definitely
wrong, i.e., it is not easier.  
Galaxy measurements really should be converted back to
observable units in the end, even if the initial massive data compression is
done in fiducial comoving units.  Given
power in observable units someone who wants to use it for broader cosmological
model fitting just needs to make predictions for it in
whatever model they want to test. Given power in fiducial comoving coordinates
they need to figure out what fiducial model was used and convert back to 
observable
coordinates before comparing (or, equivalently, apply ratios of
distances model-by-model to convert from one set of comoving coordinates to 
another). The point is not that this is very hard, it's just that there is no
ease-of-use argument for fiducial comoving coordinates. In fact, it is 
straightforward to label bands in a table with both types of coordinates, so
that, e.g., someone who wants to make a quick plot with comoving labels can
do that.

\subsection{Optimality}

If we can compute $L(\vd|\vp)$ exactly, we clearly have all the
information we can have and our analysis is therefore
optimal. 
Sub-optimality is a less well-defined concept in a Bayesian
context compared to frequentist analysis. In the latter we deal with
estimators, i.e. procedures that given data produce an estimate of the
desired quantity in which we can trade complexity of the estimator
computation with the output variance. An estimator can be suboptimal
but still unbiased. In the Bayesian analysis however, likelihood is
the basic concept and there is not such thing as ``unbiased'' likelihood.
When we Taylor expand around 
$\vp_0$ to estimate the likelihood, we are making an error in throwing out
terms beyond quadratic order. This error will generally be larger in vicinity
of the true parameters $\vp_t$ (where we would like to use it) if $\vp_0$ is 
farther away from the truth. 
As we saw in Section \ref{sec:numeric} we had to
expand $P^\times$ not around the model we would have liked to expand around, 
but
instead around a model for which we can evaluate $\vC_0^{-1}$
efficiently. 
In terms of the OQE picture above one could conclude that
our estimate is somewhat sub-optimal and our error-matrix
somewhat wrong. 
If we want to be more accurate, one way is to try
to expand around $\vp_0$ closer to $\vp_t$ 
(i.e., use a better $\vC^{-1}_0$ for weighting), but another way is to go to 
the
next order in the Taylor series. While this is rarely if ever done, it is a 
very realistic possibility given
our algorithm, because, once we have computed $\vM_{I\alpha\beta}$ for each
spectrum, the trace calculations that go into computing the Fisher matrix 
or higher derivatives are not very costly. 
The third order term could be used directly in subsequent uses of the 
results, 
or could be used to shift to a traditional Gaussian expansion around a more 
desirable model $\vp_1 \sim \vp_t$, i.e., 
\begin{eqnarray}
  \sL_{,\alpha} (\vp_1) &=&   \sL_{,\alpha} (\vp_0) +
                            \sL_{,\alpha\beta}(\vp_0)
  (\vp_1 - \vp_0)_\beta + 
\frachalf \sL_{,\alpha\beta\gamma}(\vp_0)
  (\vp_1 - \vp_0)_\beta  (\vp_1 - \vp_0)_\gamma
+\ldots \\
  \sL_{,\alpha\beta} (\vp_1) &=&   
  \sL_{,\alpha\beta} (\vp_0) + 
\sL_{,\alpha\beta\gamma}(\vp_0)
  (\vp_1 - \vp_0)_\gamma + 
\ldots,
\end{eqnarray}

In this paper, we do not go beyond second order. Our 
$\chi^2$ values on mocks suggest $\sim 8$\% underestimation of errors, 
suggesting there is not a lot of optimality to be gained by going to higher
order. It would be nice to pick up this $\sim 8$\% if it comes from 
correlations between close quasar pairs that we are not including in 
$\vC_0^{-1}$ 
\cite{2011JCAP...09..001S}, but in any case we will inevitably want some kind 
of bootstrap-like cross-check of our errors, and this should be good enough,
combined with an estimate from mocks, to convincingly calibrate a small
error in the errors.

\subsection{Speed}

In spite of being massively faster than evaluating the OQE equations in 
pixel space, our algorithm still takes longer to run in practice than the
well-developed pairwise pixel methods used to measure the \lya\ 
correlation function in BOSS 
\cite{2011JCAP...09..001S,2013A&A...552A..96B,2015A&A...574A..59D,2017A&A...603A..12B}. 
Clearly they are not evaluating the OQE equations -- they make their own set
of approximations to make their algorithm faster. Like us, they do not compute
to wide transverse separations, and do not use off-diagonal $\vC_0$ at all in 
their estimator, ignoring correlations even within pixels in the same spectra.
Moreover, most of these analyses used rebinned pixels to speed up the code, 
and measured the correlations in a single redshift bin.
We have a lot of overhead related to using a large number of parameters to 
resolve each coordinate. It is possible that this could be further reduced. 
A clear optimization which we plan as a next step is to have the number
and positions of radial (and redshift?) bins adapt to separation so
that the full shape of $P^\times$ is resolved where the information
lies and not over-resolved elsewhere. 
The slow part of our method is the calculation of the second derivative or 
equivalently the error matrix. 
Roughly half of the cost comes from pre-computing the $\vM_{I\alpha\beta}$ 
matrices and half from then computing the cross-terms between quasars.
If we extended to large angular separations, the latter would dominate.
The equivalent part of the standard correlation function analysis
would be the full evaluation of the 4-point function required to
determine the error-matrix. 
Recent standard analysis avoid this by using errors determined by data-internal
methods like bootstrap, sometimes comparing to direct calculations sped up by 
Monte Carlo subsampling of pair contributions to compute. 
We could also do this kind of subsampling.

\subsection{Application to fully realistic data}

In this paper we have ignored a vast array of systematic effects that
need to be dealt with in any real analysis. These include coaddition,
non-Gaussian resolution matrix, per spectral pixel noise estimates,
sky residuals, continuum fitting, metal contamination, etc. 
Our analysis is designed, however, to make it easier to deal with
some of them compared to standard analysis. 
We can add marginalization over any systematic template data vector 
by projecting it out during weighting. 
For instance, by adding a large constant in our quasar covariances we 
effectively project out the distortion caused by fitting the continuum
amplitude of each quasar, common in most measurements of the correlation 
function \cite{2011JCAP...09..001S,2015JCAP...11..034B,2017A&A...603A..12B}.
The marginalization increases the uncertainty in the low $k_\parallel$ 
power spectrum, but removes any bias.
We have tested that we get identical results when we substitute
$\delta \rightarrow \delta - \left< \delta \right>$, where 
$\left< \delta \right>$ is the averaged fluctuation in each line of sight.

Most metal contamination can be dealt with by subtracting the corresponding 
power
measured on the red-side of the \lya\ emission line.  
Finally, our formalism naturally allows for a unified analysis
of 1D and 3D power spectra into a single master likelihood.

\section*{Acknowledgments}
We would like to thank Simeon Bird for comments on the manuscript.
AFR acknowledges support by an STFC Ernest Rutherford Fellowship, grant reference ST/N003853/1.
AS acknowledges hospitality of the University College London.
This work was partially enabled by funding from the UCL Cosmoparticle 
Initiative.
This research used resources of the National Energy Research Scientific 
Computing Center, a DOE Office of Science User Facility supported by the 
Office of Science of the U.S. Department of Energy under Contract 
No. DE-AC02-05CH11231.

\bibliography{cosmo,cosmo_preprints}
\bibliographystyle{JHEP}

\appendix

\section{Marginalizing over unwanted parameters}
\label{sec:margin}

We start with a log-likelihood that depends on parameters that we
are interested in (labelled with greek indices $\alpha$, $\beta$) and parameters that 
we would like to marginalize over (labelled with latin indices $i$, $j$):
\begin{equation}
\sL \left( \vp \right) = \sL \left( \vp_0 \right) 
  + \sL_{,\alpha} ~ \vdelta \vp_\alpha   + \sL_{,i} ~ \vdelta \vp_i
  + \frac{1}{2}  \sL_{,\alpha\beta} ~ \vdelta \vp_\alpha ~ \vdelta \vp_\beta 
  +   \sL_{,\alpha i} ~ \vdelta \vp_\alpha ~ \vdelta  \vp_i 
  + \frac{1}{2}  \sL_{,ij} ~ \vdelta \vp_i ~ \vdelta \vp_j ~.
\end{equation}

Here we would like to calculate the marginalized likelihood
\begin{equation}
  L_m \propto \exp\left[\sL_{,\alpha} ~ \vdelta \vp_\alpha + \frac{1}{2} \sL_{,\alpha\beta} ~ \vdelta \vp_\alpha ~ \vdelta \vp_\beta \right]
  \int d^N\vp \exp\left[ \frac{1}{2}\sL_{,ij} ~ \vdelta \vp_i ~ \vdelta \vp_j  + \left (\sL_{,i} +  \sL_{,\alpha i} ~ \vdelta \vp_\alpha \right) \vdelta \vp_i  \right] ~,
\end{equation}
where the integral is over unwanted parameters
\footnote{To be accurate, we compute the integral of the posterior  
probability, i.e., we include a prior for the unwanted parameters. 
As discussed at the end of section \ref{sec:Like}, we use Gaussian priors 
around zero, with a width set to a thousand times the fiducial power 
evaluated at the center of the band.}.
To perform integration, we use the following identity:
\begin{equation}
\int_{-\infty}^{\infty} \mathcal{D}\vx \exp\left(-\frac{1}{2}\vx^t \vA \vx+
\vb^t\vx+ C\right)=
(2 \pi)^{N/2}{\rm det}\vA^{-1/2} \exp\left(\frac{\vb^t \vA^{-1} \vb}{2}+C\right)
\end{equation}
obtained by completing the square
\begin{equation}
  -\frac{1}{2} \left[ \vx^t \vA \vx - 2\vb^t\vx \right] = 
-\frac{1}{2} \left[ (\vx^T - \vA^{-1}\vb)^T \vA (\vx - \vA^{-1}\vb)
\right] +\frac{1}{2} \vb^t\vA^{-1}\vb.
\end{equation}
After integration, we get 
%% AS : just put the missing term below
% terms like $\log |\sL_{,ij}|$, but in this analysis
% we are interested in the dependence of the likelihood on the parameters 
% ($\alpha$,$\beta$), and not in its absolute value.
% \footnote{This would not be the case when computing the traditional 
% \textit{BAO scans}, where one computes the maximum of the likelihood at 
% a fixed value of the BAO scale, marginalizing over unwanted parameters. 
% In that case we would need to add a couple of extra terms.}
% We get the following marginalized likelihood:
\begin{equation}
 L_m \propto \left({\rm det} \sL_{,ij}\right)^{-1/2} \exp\left[\sL_{,\alpha} ~ \vdelta \vp_\alpha 
  + \sL_{,\alpha\beta} ~ \vdelta \vp_\alpha ~ \vdelta \vp_\beta 
  + \frac{1}{2} (-\sL^{-1}_{,ij}) \left (\sL_{,i} 
  + \sL_{,\alpha i} ~ \vdelta \vp_\alpha \right) \left (\sL_{,j} 
  + \sL_{,\beta j} ~ \vdelta \vp_\beta \right) \right] ~.
\end{equation}
Therefore, we can compute the marginalized first and second derivatives:
\begin{eqnarray}
 \sL_{m,\alpha} &=& \sL_{,\alpha} 
    + (-\sL^{-1}_{,ij}) \left (\sL_{,\alpha i} ~ \sL_{,j}\right) \\
 \sL_{m,\alpha\beta} &=& \sL_{,\alpha\beta} 
    + (-\sL^{-1}_{,ij}) \left (\sL_{,\alpha i} ~ \sL_{,\beta j}\right) ~.
\end{eqnarray}

Note that in the equations above we need to invert the submatrix of second 
derivatives with respect to unwanted parameters, $\sL_{,ij}$. 
If the likelihood is not sensitive to one of these parameters, this matrix
will be non-positive definite.  
When needed, we regularise those direction in parameter space by adding a 
small negative value to the diagonal elements $\sL_{,ii}$.

\section{3D power spectrum window function}
\label{sec:derivatives}

In section \ref{sec:P3D} we have motivated a particular form of 
$\frac{\partial \px_\alpha }{\partial \ptD_\gamma}$:
\begin{equation}
\frac{\partial p^\times_\alpha}{\partial p^{3D}_\gamma} =
  \int \frac{d\vk_\perp}{(2\pi)^2} ~ e^{i \vDtheta_\alpha \vk_\perp}
  w_\gamma(z_\alpha,\vk_\perp,\kpar^\alpha) 
\label{eq:cr2}
\end{equation}
This form is exact in the limit of infinite number of (intermediate)
$\px$ and $\ptD$ parameters. 
Given that our results do not vary if we use a finer grid of parameters, we 
believe that this form is accurate enough for our analysis.

If we wanted to use a coarser parameterization, we would have to compute an
effective window function.
Even though this is not done in our analysis, in this appendix we describe 
how it could be done.

Equation \ref{eq:cr2} implies that $p^\times$ parameters are modeled by
\begin{equation}
 P^{\times \rm fid}_\times(\vDtheta_\alpha)+ p^\times_\alpha = 
  P_{\times}^{3D{\rm fid}}(\vDtheta_\alpha) + \sum_\gamma p_\gamma^{3D} 
  \frac{\partial p^\times_\alpha}{\partial p^{3D}_\delta},
\end{equation}
where $P^{\times \rm fid}_\times$ is the fiducial cross-power spectrum, and 
$P^{3D \rm fid}_\times$ is the cross-power spectrum computed from the 
fiducial 3D power spectrum.
The equivalent of equation \ref{eq:inverting} is therefore
\begin{equation}
p^{3D}_\gamma = \sum_\delta J^{-1}_{\gamma\delta} \sum_\alpha \left[
 p^\times_\alpha + P_\times^{\times \rm fid} (\vDtheta_\alpha)  
 - P_\times^{3D \rm fid}(\vDtheta_\alpha) \right]
 \frac{\partial p^\times_\alpha}{\partial p^{3D}_\gamma} ~,
 \label{eq:p3D19}
\end{equation}
with
\begin{equation}
 J_{\gamma\delta} = \sum_\alpha 
  \frac{\partial p^\times_\alpha}{\partial p^{3D}_\gamma} 
  \frac{\partial p^\times_\alpha}{\partial p^{3D}_\delta}.
\end{equation}

We also know that
\begin{equation}
 \px_\alpha = I^{-1}_{\alpha\beta} \int d^2 \vDtheta w^\times_\beta(\Dtheta) 
  \left(P_\times (\Dtheta) - P^{\rm fid}_\times(\Dtheta) \right) ~,
 \label{eq:px19}
\end{equation}
where 
\begin{equation}
 I_{\alpha\beta} = \int d^2 \vDtheta w^\times_\alpha(\Dtheta) 
  w^\times_\beta(\Dtheta)  
\end{equation}
and $w^\times$ are the cross-power interpolation kernels in angular
direction.\footnote{
The above equations are trivially generalizable to  the case with non-uniform
error bars using (Einstein convention assumed):
\begin{eqnarray}
 p^{3D}_\gamma &=&  J^{-1}_{\gamma\delta}  
    \left[p^\times_\alpha - P_{\times}^{3D{\rm fid}}(\vDtheta_\alpha) \right]
    C^{\times -1}_{\alpha\beta}
    \frac{\partial p^\times_\beta}{\partial p^{3D}_\gamma} ~,\\
 J_{\gamma\delta} &=& \frac{\partial
    p^\times_\alpha}{\partial p^{3D}_\gamma} C^{\times -1}_{\alpha\beta}
    \frac{\partial p^\times_\beta}{\partial p^{3D}_\delta}.
\end{eqnarray}
and non-uniform density of skewers given by $f(\vDtheta)$:
\begin{eqnarray}
 \px_\alpha &=& I^{-1}_{\alpha\beta} \int d^2 \vDtheta 
    w^\times_\beta(\Dtheta) f(\Dtheta) 
    \left(P_\times (\Dtheta) - P^{\rm fid}_\times(\Dtheta) \right) ~,\\
 I_{\alpha\beta} &=& \int d^2 \vDtheta  w^\times_\alpha(\Dtheta)
    w^\times_\beta(\Dtheta) f(\Dtheta)   ~.
\end{eqnarray}
}

We can now put together equations \ref{eq:p3D19} and \ref{eq:px19} to get:
\begin{equation}
  \ptD_\gamma=\sum_{\delta,\alpha,\beta} J^{-1}_{\gamma \delta}
  I^{-1}_{\alpha \beta} \iiint d^2 \vDtheta d^2 \kperp d^2\kperp' 
  \left(P^{3D}(\kperp,\kpar,z) - P^{3D{\rm fid}}(\kperp,\kpar,z)
  \right) e^{i(\Delta \theta \kperp + \Delta \theta_\alpha \kperp')}
    w^\times_\beta(\Delta \theta) w_\delta^{3D}(\kperp')
\end{equation}
The above provide the prescription on how to calculate the window
function in the $\kperp$ direction and can be simplified significantly
by inserting concrete forms for windows $w$. The response in $\kpar$
and $z$ direction is given simply by Equation \ref{eq:inverting}.
For sufficiently narrow bins, $\ptD_\gamma =
P^{3D}(k_{\perp\gamma},\kpar,z)- P^{3D{\rm fid}}(k_{\perp\gamma},\kpar,z)$.

\end{document}